\documentstyle[aps, epsf, amsmath, fleqn]{revtex}

\def \lam{\lambda}
\def \half{\frac{1}{2}}
\def \eps{\epsilon}
\def \d{\partial}
\def \calt{{\cal\char'124}}
\def \mz{\widehat{\cal\char'131}}

\def \non{\nonumber}

\renewcommand{\(}{\left(}
\renewcommand{\)}{\right)}
\newcommand{\df}[2]{ \frac{\partial {#1}}{\partial {#2}} }
\newcommand{\dff}[2]{ \frac{\partial^2 {#1}}{\partial {#2}^2} }

\newcommand{\wh}{\widehat}

\begin{document}

\title{Excitation of Neutron Star Oscillations by an Orbiting Particle}

\author{Johannes Ruoff${}^{(1,2)}$,  
        Pablo Laguna${}^{(2)}$ and Jorge Pullin${}^{(2)}$}

\address{${}^{(1)}$ Institut f\"ur Astronomie und Astrophysik\\ Universit\"at
T\"ubingen, D-72076 T\"ubingen, Germany}

\address{${}^{(2)}$ Center for Gravitational Physics \& Geometry\\
Penn State University, University Park, PA 16802, USA}

\maketitle

\begin{abstract}
The excitation of neutron stars is expected to be an important source
of gravitational radiation. Of fundamental importance is then to
investigate mechanisms that trigger oscillations in neutron stars in
order to characterize the emitted radiation. We present results from a
numerical study of the even-parity gravitational radiation generated
from a particle orbiting a neutron star. We focus our investigation
on those conditions on the orbital parameters that favor the excitation of
$w$-modes. We find that, for astrophysically realistic conditions,
there is practically no $w$-mode contribution to the emitted radiation. 
Only for particles with ultra-relativistic orbital speeds $\ge 0.9c$,
the $w$-mode does significantly contribute to the total emitted
gravitational energy. We also stress the importance of setting
consistent initial configurations.
\end{abstract}

\widetext

\section{Introduction}
\label{sec:introduction}
Neutron stars as a source of gravitational radiation have been under
investigation for over 30 years, with the foundation developed by the work
of Thorne and Campolattaro \cite{TC67}. The radiation emitted by an excited
non-rotating neutron star basically consists of a superposition of the
characteristic oscillatory modes of the star \cite{Koj88,KS92,LNS93}.
Those modes fall into two categories. One class are spacetime modes, such
as the $w$-modes and the trapped modes for very compact star. The other class
consist of fluid modes, such as the $f$-mode and the $p$-modes which also exist
in Newtonian stars. Due to the emission of gravitational waves, both types
of modes are damped, with the spacetime modes much stronger than the
fluid modes.

Recent studies of the time evolution of these oscillatory modes
\cite{AAKS98,Ruoff00} have shown that generic, but somewhat
unphysical, initial data can excite the first $w$-mode, the $f$-mode,
and a few $p$-modes. However, the strength of these modes depends
crucially on the particular choice of initial data. Conformally flat
initial data, for instance, has the effect of almost completely
suppressing the excitation of $w$-modes \cite{AAKLPR99}, whereas the
$f$-mode is generally present. Work by Andersson and Kokkotas
\cite{AK98} shows that, by extracting the frequencies and damping times
of the first $w$-mode and the $f$-mode, one could in principle
determine physical parameters of the neutron star such as mass,
radius and even the equation of state. Of course, the success in the
determination of such physical parameters will depend on the
particular modes present in the signal and the noise level.

The issue of interest for the work in this paper is whether
astrophysically motivated initial neutron star perturbations are able
to excite the $w$-modes, so that accurate determination
of the mode parameters could be possible. One of the most likely
events that may excite strong oscillations is the birth of the neutron
star. However, a study of this process would require a rather complete
understanding of the collapse of the progenitor star within the full
nonlinear theory of general relativity.
The alternative is to study astrophysical systems involving neutron
star oscillations that can be treated using perturbation theory. A
system that allows such an approach and, at the same time, could be
viewed as an approximation to a realistic astrophysical system is that
of a ``small'' particle orbiting a neutron star. Here small is to be
understood in the sense that the mass $\mu$ and size of the particle
are much smaller than the mass $M$ and radius $R_*$ of the neutron
star. Under this approximation, the particle moves along a geodesic in
the spacetime generated by the neutron star, causing metric
perturbations that propagate and eventually excite the neutron
star. Of course, in principle those gravitational perturbations in
turn would cause deviations of the particle's trajectory; however,
these back-reaction effects are of second order and will therefore be
neglected.

The specific goal of this paper is to investigate the gravitational
radiation emitted from the orbiting and scattering of a particle by a
neutron star. Collisions of particles with the neutron star
\cite{Bor97} are not considered because it is not clear at all how to
treat the impact and subsequent merge of the particle with the neutron
star. Contrary to previous studies of excitations of neutron stars by
orbiting particles, we perform our calculations in the time domain
instead of the frequency domain \cite{FGB99,TS99,AP99}. In particular,
we use the linearized ADM-formalism \cite{ADM62}, which is naturally
adapted for the numerical initial value problem. The relevant
equations have been derived in \cite{Ruoff00}, which we shall refer to
as paper I. Considering the dynamics in the time
domain has as a consequence that we are forced to ``smooth out'' the
particle; that is, the $\delta$-functions in the sources of the
equations due to the presence of the particle are approximated by
Gaussians. We show, however, that our model of the particle is
self-consistent and convergent. Another important issue is the
prescription of appropriate initial data that satisfy the constraints.

The paper is organized as follows. In section \ref{sec:model}, we
present the equations to describe the unperturbed configuration and
its perturbations, as well as the method to treat the orbiting
particle as a perturbation of the background spacetime. Section
\ref{sec:num} deals with the numerical approximation of the particle
terms in the equations. In section \ref{sec:init}, we outline the
problems with choosing appropriate initial data, and, by referring to
the flat space problem, we give an approximate prescription of
``good'' initial data. In sections \ref{sec:res} and \ref{sec:con}, we
present and discuss our numerical results. Details concerning the
derivation the particle terms can be found in the appendix.

\section{Formulation of the problem}
\label{sec:model}

The energy-momentum tensor $\calt^{\mu\nu}$ of a point particle with
mass $\mu$ is given by
\begin{align}
	\calt^{\mu\nu} &= \mu\int\delta^{(4)}\(x^\kappa-X^\kappa(\tau)\)u^\mu
	u^\nu d\tau\non \\
	&= \mu\frac{U^\mu U^\nu}{U^t r^2}
	\delta(r-R(t))\delta\(\phi - \Phi(t)\)
	\delta\(\cos\theta - \cos\Theta(t)\)\;,
\label{part_tmn}
\end{align}
where $U^\mu$ is the 4-velocity of the particle
\begin{align}
        U^\mu &= \frac{dX^\mu}{d\tau}\;,
\end{align}
and $\tau$ the particle's proper time along its trajectory $X^\kappa(\tau)$. 

We now orient the coordinate system in such a way that the
particle's orbit coincides with the equatorial plain of the neutron
star ($\Theta = \pi/2$). Besides, the particle's coordinate time
$T$ is identical with the time coordinate $t$ of the spacetime in
which it moves. Therefore we will use $t$ to parametrize the
path of the particle $X^\mu(t) = [t,\, R(t),\,\pi/2,\,\Phi(t)]$.

From the geodesic equations $\frac{DU^\mu}{d\tau} \equiv U^\nu
U^\mu_{\phantom{\mu};\nu} = 0$ we find:
\begin{subequations}\label{geo_eqs}
\begin{align}
	\label{geo1}
	\frac{dt}{d\tau} &= e^{2\lambda}E\\
	\label{geo2}
	\frac{dR}{d\tau} &= \left [E^2 - e^{2\nu}\(1 + \frac{L^2}{R^2}\)\right ]^{1/2}\\
	\label{geo3}
	\frac{d\Phi}{d\tau} &= \frac{L}{R^2}\;,
\end{align}
\end{subequations}
where $E$ and $L$ are the energy and angular momentum per unit mass of
the particle, respectively. We also recall that for the Schwarzschild
metric we have
\begin{align}
	e^{2\nu} &= e^{-2\lambda} \;=\; 1 - \frac{2M}{r}\;.
\end{align}
We can use \eqref{geo1} to eliminate
the proper time $\tau$ from equations \eqref{geo2} and \eqref{geo3}:
\begin{subequations}
\begin{align}
	\label{geo2t}
	\frac{dR}{dt} &= e^{2\nu}\left [1
	- \frac{e^{2\nu}}{E^2}\(1 + \frac{L^2}{R^2}\)\right ]^{1/2}\\
	\label{geo3t}
	\frac{d\Phi}{dt} &= e^{2\nu}\frac{L}{R^2E}\;.
\end{align}
\end{subequations}
Eqs.~\eqref{geo2t} and \eqref{geo3t} can be used to replace the
quantities $v_r \equiv \frac{dR}{dt}$ and $v_\phi \equiv
\frac{d\Phi}{dt}$ in the source terms of the particle. However, we
also have to explicitly solve them for the particle's trajectory
coordinates $R(t)$ and $\Phi(t)$, since we need those coordinates in
the $\delta$-functions $\delta(r-R(t))$ and $\delta(\phi-\Phi(t))$.

The background spacetime is solely due to the neutron star, which we
model as a non-rotating polytrope with $\Gamma = 2$ and $\kappa =
100\,$km$^2$. Outside the star, the spacetime is given by the
Schwarzschild metric. The equations governing the stellar oscillations
are derived in paper I. After expansion in Regge-Wheeler tensor
harmonics, these equations can be grouped into two families, one describing the
axial perturbations and the other describing the polar perturbations.

In the exterior region, we now have to incorporate the source terms due
to the particle into the perturbation equations. For that purpose, the
energy-momentum tensor of the particle has to be expanded in tensor
harmonics. The derivation of this expansion is given 
in the Appendix. There is a slight
inconvenience created by the fact that the perturbation equations are
expanded using the Regge-Wheeler harmonics \cite{RW57}, which do not
form an orthonormal set, whereas the energy-momentum tensor of the
particle is expanded using the orthonormal harmonics given by Zerilli
\cite{Zer70a}. However, the conversion is straightforward.

Even though the main focus will be on the polar perturbations, we
also present the axial equations since they are so simple. In the
Regge-Wheeler gauge, the only non-vanishing axial perturbations of the
shift vector and spatial metric 
are respectively given by
\begin{subequations}
\begin{align}
	\(\beta_\theta,\,\beta_\phi\) &= e^{\nu-\lam}K_6^{lm}
	\(-\sin^{-1}\theta\df{}{\phi}Y_{lm},\,
	\sin\theta\df{}{\theta}Y_{lm}\)\\
	\(h_{r\theta},\,h_{r\phi}\) &= e^{\lam-\nu}V_4^{lm}
	\(-\sin^{-1}\theta\df{}{\phi}Y_{lm},\,
	\sin\theta\df{}{\theta}Y_{lm}\)\;.
\end{align}
\end{subequations}
Similarly, the extrinsic curvature perturbations read
\begin{subequations}
\begin{align}
	\(k_{r\theta},\,k_{r\phi}\) &= \half e^{\lam}K_3^{lm}
	\(-\sin^{-1}\theta\df{}{\phi}Y_{lm},\,
	\sin\theta\df{}{\theta}Y_{lm}\)\\
	\(\begin{array}{cc}
	k_{\theta\theta} & k_{\theta\phi}\\
	k_{\phi\theta} & k_{\phi\phi}\end{array}\) &= e^{-\lam}K_6^{lm}
	\(\begin{array}{cc}
	-\sin^{-1}\theta\,X_{lm} & \sin\theta\,W_{lm}\\
	\sin\theta\,W_{lm} & \sin\theta\,X_{lm}\end{array}\)\;,
\end{align}
\end{subequations}
with $W_{lm}$ and $X_{lm}$ defined in Eqs.~\eqref{Wdef} and
\eqref{Xdef} of the Appendix. With the inclusion of the particle terms
the evolution equations read (see Paper I)
\begin{align}\label{odd1}
	\df{V_4}{t} &= e^{4\nu}\left [\df{K_6}{r}
	 + 2\(\nu' - \frac{1}{r}\)K_6\right ] - e^{2\nu}\,K_3\\
	\df{K_3}{t} &= \frac{l(l+1) - 2}{r^2}\,V_4
	- 16\pi\,e^{2\nu}t_7\\
	\df{K_6}{t} &= \df{V_4}{r} - 8\pi\,t_{10}\;.
\end{align}
The momentum constraint relates the extrinsic curvature coefficients
to the particle's source term via
\begin{align}\label{odd2}
	\df{K_3}{r} + \frac{2}{r}\,K_3
	- \frac{l(l+1) - 2}{r^2}\,K_6
	&= 16\pi\,e^{2\lambda}\,t_4\;.
\end{align}
For completeness, we also list the source terms of the particle:
\begin{align}
	t_4 &= e^{2\nu}\frac{\mu L}{r^2 l\(l+1\)}\,\delta(r-R(t))
	\df{}{\Theta}Y_{lm}^*(\frac{\pi}{2},\Phi(t))\\
	\label{t7}
	t_7 &= e^{2\lambda}\frac{\mu L}{r^2 l\(l+1\)}\,v_r\,
	\delta(r-R(t))\df{}{\Theta}Y_{lm}^*(\frac{\pi}{2},\Phi(t))\\
	\label{t10}
	t_{10} &= -e^{2\nu}\frac{2\mbox{i}m\mu L^2}
	{r^2 E\,l\(l+1\)\(l-1\)\(l+2\)}\,\delta(r-R(t))
	\df{}{\Theta}Y_{lm}^*(\frac{\pi}{2},\Phi(t))\;.
\end{align}
For radial infall, $L = 0$, and all source terms vanish. Hence, in
this case the radiation is of pure even parity.

Let us now turn to the polar perturbations. Here, we have to use some
caution in including the particle terms. From the ADM split, we see
that the matter terms enter only in the equations for the extrinsic
curvature coefficients $\wh K_i$. In the Regge-Wheeler gauge, we are
only left with equations for the coefficients $\wh K_1$, $\wh K_2$ and
$\wh K_5$. In paper I, we showed that we have to choose initial data
with $\wh K_4 = 0$. Without the particle, the evolution equations then
guaranteed the vanishing of $\wh K_4$ for all times. However, in the
presence of the particle, this is not true anymore since $\wh K_4$
satisfies the following evolution equation:
\begin{align}\label{k4}
	\df{}{t}\wh{K}_4 &= -8\pi\,t_8\;.
\end{align}
This means that during the evolution $\wh K_4$ will be become nonzero
even if it was set to zero initially. A non-vanishing $\wh K_4$
implies nonzero metric components $\wh{T}_1$ and $\wh{V}_3$; this
violates the Regge-Wheeler gauge condition. A solution is to choose a
different lapse function perturbation $\alpha$. If we pick
\begin{align}
	\alpha &= -\half\,e^{\nu}\(\frac{T}{r} + rS
	+ 16\pi\,t_8\)Y_{lm}\;,
\end{align}
the last term exactly cancels the right-hand side of (\ref{k4}) and the
vanishing of $\wh{K}_4$ and therewith $\wh{T}_1$ and $\wh{V}_3$ is,
again, guaranteed.

The remaining metric quantities are expanded in the same way as in
paper I:
\begin{align}
	\beta_i &= \(e^{2\lambda}K_2,\,0,\,0\)Y_{lm}\\
	h_{ij} &= \label{hij_polar}
	\(\begin{array}{ccc}
	e^{2\lambda}\(\frac{T}{r} + rS\) & 0 & 0\\
	0 & rT & 0\\
	0 & 0 & r\sin^2 T
	\end{array}\)Y_{lm}\;,
\end{align}
and the expansion of the extrinsic curvature reads
\begin{align}
	k_{ij} &= 
	-\frac{e^{-\nu}}{2r}\(\begin{array}{ccc}
	e^{2\lambda}K_1 & -e^{2\lambda}K_2\df{}{\theta}& 
	-e^{2\lambda}K_2\df{}{\phi}\\
	\star & r^2\(K_5 - 2K_2\)& 0\\
	\star & 0 & r^2\sin^2\theta \(K_5 - 2K_2\)
	\end{array}\)Y_{lm}\;.
\end{align}
In paper I, we have shown that it is possible to write the evolution
equations as a set of two coupled wave equations only in terms of the
metric components $S$ and $T$. With the inclusion of the particle
terms those evolution equations read in the exterior region
\begin{align}
\begin{split}
        \label{S}
        \dff{S}{t} &= e^{4\nu}\bigg[
        \dff{S}{r} + 6\nu'\df{S}{r} + \(4(\nu')^2 - 2\frac{\nu'}{r}
        - e^{2\lambda}\frac{l(l+1)}{r^2}\)S
        + 4\frac{\nu'}{r^2}\(\nu' - \frac{3}{r}\)T
	+ \frac{16\pi}{r}\,P_1\bigg]
\end{split}\\
       \label{K5}
        \dff{T}{t} &= e^{4\nu}\bigg[
        \dff{T}{r} + 2\nu'\df{T}{r} + \(2\frac{\nu'}{r}
	- e^{2\lambda}\,\frac{l(l+1)}{r^2}\)T - 2\,S
	+ 16\pi\,P_2\bigg]\;,
\end{align}
where the $P_1$ and $P_2$ denote the source terms of the 
particle
\begin{align}
\begin{split}
	P_1 &= t_8'' - 2\,t_6' + \(5\,\nu' - \frac{3}{r}\)t_8'
	+ t_5 - 2\(3\,\nu' - \frac{1}{r}\)t_6
	+ 2\(\nu'\(\nu' - \frac{6}{r}\) + \frac{2}{r^2}\)t_8
	- \frac{e^{2\lambda}}{r^2}\,t_9
\end{split}\\
	P_2 &= t_8' - 2\,t_6
	+ 2\(\nu' - \frac{1}{r}\)t_8 + \frac{e^{2\lambda}}{r}\,t_9
	- \frac{r}{2}\,e^{2\lambda}\,t\;,
\end{align}
with the $t_i$ given in the Appendix. 
The evolution equations which govern the oscillations in the interior
of the neutron star are given in paper I (e.g.~(49) and (58)).
Furthermore, the perturbations in the exterior are completely
described by the Zerilli equation (eqs. (61) of paper I):
\begin{align}\label{Zeqn}
	 \dff{Z}{t} &= \frac{\d^2Z}{\d r^2_*} + V(r)Z + S_Z\;,
\end{align}
with the following explicit form of the source term
\begin{align}
\begin{split}
	S_Z &= -16\pi e^{4\nu}\frac{\mu}{r^2E\Lambda(n+1)}
	\bigg[e^{2\nu}(L^2 + r^2)\delta'(r-R) + \bigg(
	3M\(1 + 4\frac{E^2}{\Lambda}\) - r(n+1) - 2\mbox{i}m e^{2\mu}v_rLE\\
	& \hspace*{4cm} + \frac{L^2}{r^2n}
	\(rn(m^2 - 3 - 2n) - M(3-2n - 3m^2)\)\bigg)\delta(r-R)\bigg]Y^*_{lm}\;,
\end{split}
\end{align}
where $2n = l(l+1) - 2.$
In addition, we have three constraint equations, namely the
Hamiltonian constraint, which now reads
\begin{align}\label{hc}
        T'' + \nu'T' - rS'&
        - \(5\frac{\nu'}{r} + e^{2\lambda}\frac{l(l+1)}{r^2}\)T
        - \(2 r\nu' + 2 + \half e^{2\lambda}l(l+1)\)S
	&= -8\pi e^{2\lambda}\frac{\mu E}{r}\,\delta(r-R(t))
	Y_{lm}^*\;,
\end{align}
and two momentum constraints
\begin{align}\label{mc1}
	rK_5' - \half e^{2\lambda}\,l(l+1)\,K_2
	- r^2K - \(r\nu' + 1\)K_5
	&= 8\pi e^{2\mu}\mu E
	v_r\,\delta(r-R(t))Y_{lm}^*\\
	rK_2' - r^2K - 2\,K_5\label{mc2}
	&= 16\pi e^{2\nu}\frac{\mbox{i}m\mu L}{r l\(l+1\)}\delta(r-R(t))
	Y_{lm}^*\;.
\end{align}
Notice that the quantity $K_2$ appears in Eqs.~\eqref{mc1} and
\eqref{mc2} but not in the evolution equations. Therefore, we will
eliminate it by differentiating (\ref{mc1}) with respect to $r$ and
using (\ref{mc1}) and (\ref{mc2}) to eliminate $K_2$ and $K_2'$. The
resulting equation is then second order in $K_5$ and reads
\begin{align}
\begin{split}\label{MC2nd}
	K_5'' + \nu'K_5'  - rK'& - \(5\frac{\nu'}{r}
	+ e^{2\lambda}\frac{l(l+1)}{r^2}\)K_5
	 - \(2r\nu' + 2 + \half e^{2\lambda}l(l+1)\)K\\
	&= 8\pi\frac{\mu}{r}\(e^{2\lambda}v_rE\delta'(r-R(t))
	+ \mbox{i}m L\delta(r-R(t))\)Y_{lm}^*\;.
\end{split}
\end{align}
It should be noted, that this equation also follows from taking the
time derivative of the Hamiltonian constraint (\ref{hc}) and taking
into account that $\dot{S} = K$ and $\dot{T} = K_5$.

As usual, the initial data must satisfy the constraint equations.
Unfortunately, there is no unique way to solve those
equations. This is due to the fact that to a particular solution of
the inhomogeneous equations, we can always add a solution of the
homogeneous equation, which would correspond to adding some arbitrary
gravitational waves. The problem of finding the ``right'' initial data
that represent only the perturbations which are due to the presence of
the particle and which do not contain any additional radiation will be
discussed in more detail in section \ref{sec:init}.

For radial infall of a particle from rest, $v_r(t = 0) = 0$ and $L =
0$, and therefore $t_2 = t_3 = 0$. Thus the momentum constraints
(\ref{mc1}) and (\ref{mc2}) can be trivially satisfied by setting the
extrinsic curvature variables to zero. This situation corresponds to time
symmetric initial data. We then are left with solving the Hamiltonian
constraint (\ref{hc}). Of course, a particle falling from rest would
fall radially towards the neutron star and eventually hit its surface.
Since we want to avoid such an impact, we have to give the particle
some angular momentum. In addition, we want to consider different
initial radial velocities which means that we have to solve also
the momentum constraints (\ref{mc1}) and (\ref{mc2}).

\section{Numerical implementation of the particle}
\label{sec:num}

The presence of a particle in the calculations introduces source terms
in the equations that require the 
explicit forms of the spherical harmonics $Y_{lm}$. The perturbation
equations without particle are degenerate with respect to $m$ since
the background metric is spherically symmetric. However, 
the particle breaks this symmetry, and one is forced to consider the
various $m$--cases. Fortunately, we do not have to consider all
possible values of $m$ for a given values of $l$ since for negative
$m$ the spherical harmonics just undergo sign change and phase shift
($Y^*_{lm} = (-1)^m\, Y_{l,-m}$). The advantage of putting the
particle in the equatorial plane ($\Theta = \frac{\pi}{2}$) is that, in
the even parity case, we only have to deal with multipoles with $m = l$,
$l-2$, ... , the remaining ones ($m = l-1$, $l-3$, ...) have odd
parity. Since the evolution code handles only real valued
perturbations, we have to treat the real and imaginary parts of the
spherical harmonics separately. Finally, all the equations will be
solved on a finite grid, hence we have to approximate the
$\delta$--function by a narrow Gaussian
\begin{align}
        \delta(r-R(t)) &\approx \frac{1}{\sigma\sqrt{2\pi}}\,
        e^{-\frac{\(r - R(t)\)^2}{2\sigma^2}}\qquad\mbox{$\sigma$ small}\;.
\end{align}
To demonstrate the validity of this approximation, we have to ensure the
convergence of the solution for $\sigma \rightarrow 0$. This can be done
in two different ways. One can look at the convergence of the
waveforms that are obtained in the evolution, or also one 
can monitor the violation of the constraints. A possible way do to so is
to monitor the following quantity
\begin{align}\label{I}
	I &= \frac{1}{8\pi\,\mu E\,Y^*_{lm}}\,\int{r e^{2\nu}
	\(\mbox{lhs of (\ref{hc})}\)dr}\;,
\end{align}
where the domain of integration is the region outside the neutron
star. In the limit $\sigma\rightarrow 0$, one gets that $I = 1$
throughout the whole evolution. Numerically, we cannot take this limit
with a fixed grid size since, eventually, we cannot sufficiently
resolve the Gaussian. To check the convergence, we decrease both
$\sigma$ and $\Delta x$ by keeping the ratio $\sigma/\Delta x$
constant throughout the sequence. By consecutively doubling the
resolution and halving $\sigma$, we find that $I - 1$ approaches zero
with second order convergence. For the numerical evolutions, we will
use a value of $\sigma/\Delta x = 0.15$, which gives provides a
suitable resolution of the Gaussian and its derivatives.

There is still a subtle point. In deriving the source terms (see
Appendix), we have tacitly transformed the particle coordinate $R$
into the spacetime coordinate $r$ since the presence of the
$\delta$-function makes a distinction unnecessary. However, in the
evolution equations we have to take derivatives of the source terms
with respect to $r$ but not with respect to $R$, and therefore we
would obtain different source terms if we had not changed the $R$'s
into $r$'s. As an example consider the following two source terms
\begin{align}
	S_1(r) &= f(r)\delta(r - R)
\end{align}
and
\begin{align}
	S_2(r) &= f(R)\delta(r - R)\;,
\end{align}
which are equivalent because of the presence of the $\delta$-function.
But if we now differentiate $S_1$ and $S_2$ with respect to $r$ we
obtain for $S_2$ just the derivative of the $\delta$-function, whereas
for $S_1$ we also have to differentiate $f$. Analytically this does
not make a difference, but if we approximate the $\delta$-function by
a Gaussian, then the two expressions for $S_1$ and $S_2$ and their
respective derivatives are different. To gain accuracy, we
should have kept $r^2d\phi/d\tau\delta(r - R)$ as
$Lr^2/R^2\delta(r - R)$ and not just as $L\delta(r - R)$. However, for
the numerical evolutions the actual difference is negligible, so 
we have assumed the source terms to be of the form
of $S_1$.

\section{Setting up the initial conditions}
\label{sec:init}
\begin{figure}[hhh]
\leavevmode
\begin{minipage}{8.5cm}
\epsfxsize=\textwidth
\epsfbox{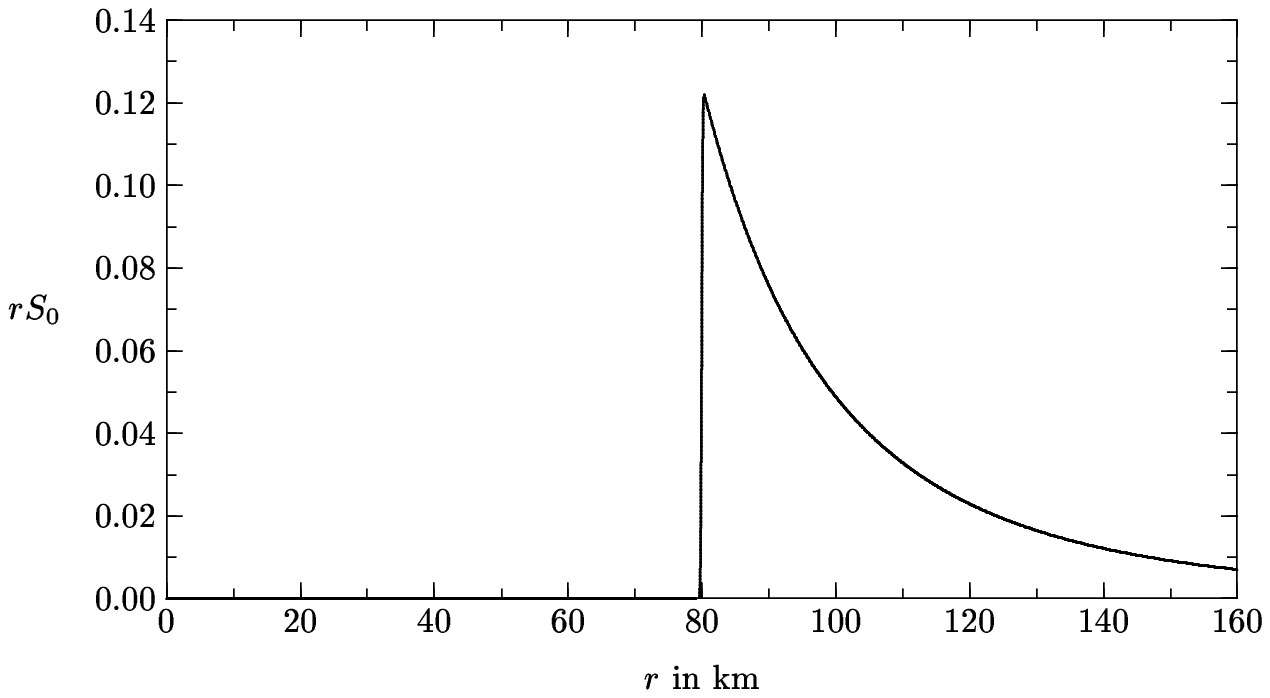}
\end{minipage}
\begin{minipage}{8.8cm}
\vspace*{1mm}
\epsfxsize=\textwidth
\epsfbox{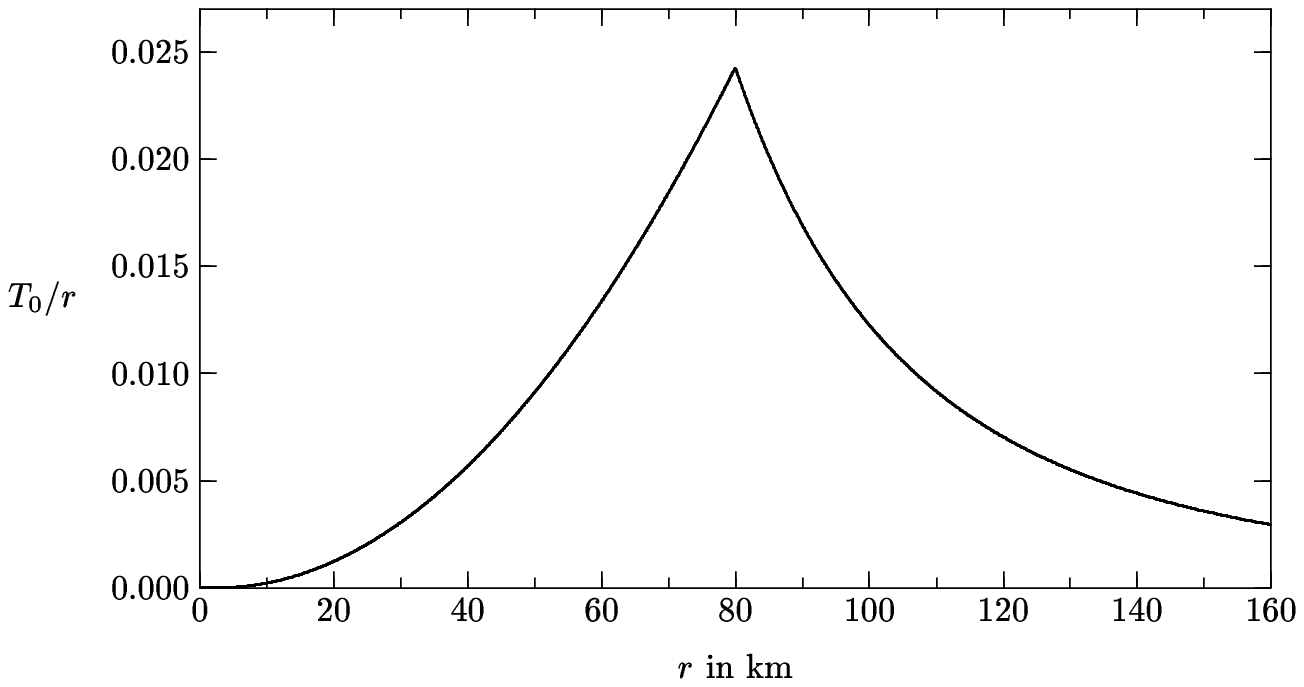}
\end{minipage}
\vspace*{3mm}
\caption{\label{init}Profile of the initial values for both kind of initial
	data. The left panel shows $S_0$ when $T_0 = 0$. The right
	panel shows $T_0$ when $S_0 = 0$. Note, that $S$ exhibits
	a discontinuity at the particle's location whereas $T$ is continuous.}
\end{figure}

Any construction of initial data for a particle initially located at
($r_0$, $\Phi_0$) with initial radial velocity $v_r$ and angular
momentum $L$ involves solving the Hamiltonian constraint (\ref{hc}).
Usually, we set the initial angle $\Phi_0$ to zero. However,
Eq.~(\ref{hc}) contains two quantities $S$ and $T$. This 
means that there is some freedom in choosing the initial values.  We
can e.g.~either set the quantity $T$ to zero and solve for $S$, or do
it the other way round and set $S = 0$ and solve for $T$. In the
former case, we would have to solve a first order equation for $S$, and
the latter case would lead to a second order equation for $T$.
In Fig.~\ref{init} we show both possible sets of initial data.

To assess which choice is the more natural, we consider a particle
initially at rest in flat spacetime. Of course, the particle will
remain at rest since there is no matter around that could attract the
particle. Thus, the perturbation of the spacetime that is created by the
particle will be stationary. Hence the equations of motion for the
metric perturbations $S$ and $T$ will read $\df{S}{t} = 0$ and
$\df{T}{t} = 0$. From these conditions and from the Hamiltonian
constraint, it then follows that $S$ has to vanish.

Of course, in the presence of the neutron star, those arguments do not
hold any more, and $S\equiv0$ will not be the right choice of initial
data, but if the particle is initially far enough away from the
neutron star, the error in setting $S=0$ should be very small. This
error actually corresponds to an introduction of an extra amount of
gravitational radiation that is not at all related to the radiation
that is emitted when the particle moves through the spacetime. This
extra amount will start to propagate during the evolution and
eventually hit the neutron star and thereby cause it to oscillate.
However, if we put the particle far enough away from the neutron star,
the strength of the induced oscillation should be small compared to
the ones excited when the particle comes close to the neutron star.

\begin{figure}[hhh]
\leavevmode
\hspace*{-1cm}
\begin{minipage}{10cm}
\epsfxsize=\textwidth
\epsfbox{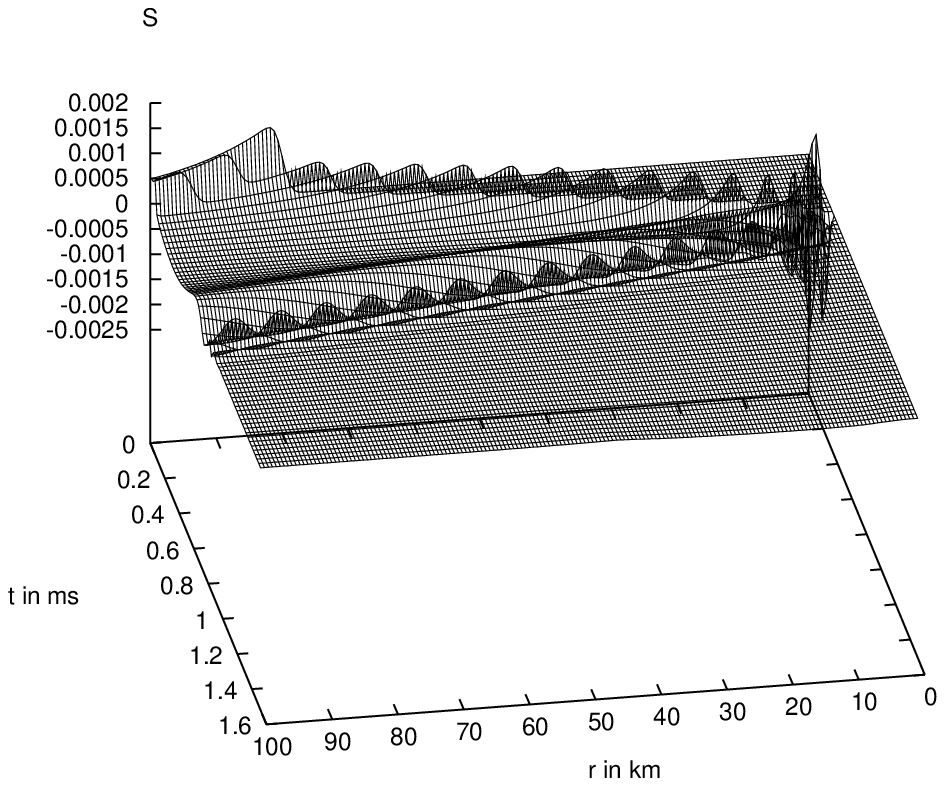}
\end{minipage}
\hspace*{-1.8cm}
\begin{minipage}{10cm}
\epsfxsize=\textwidth
\epsfbox{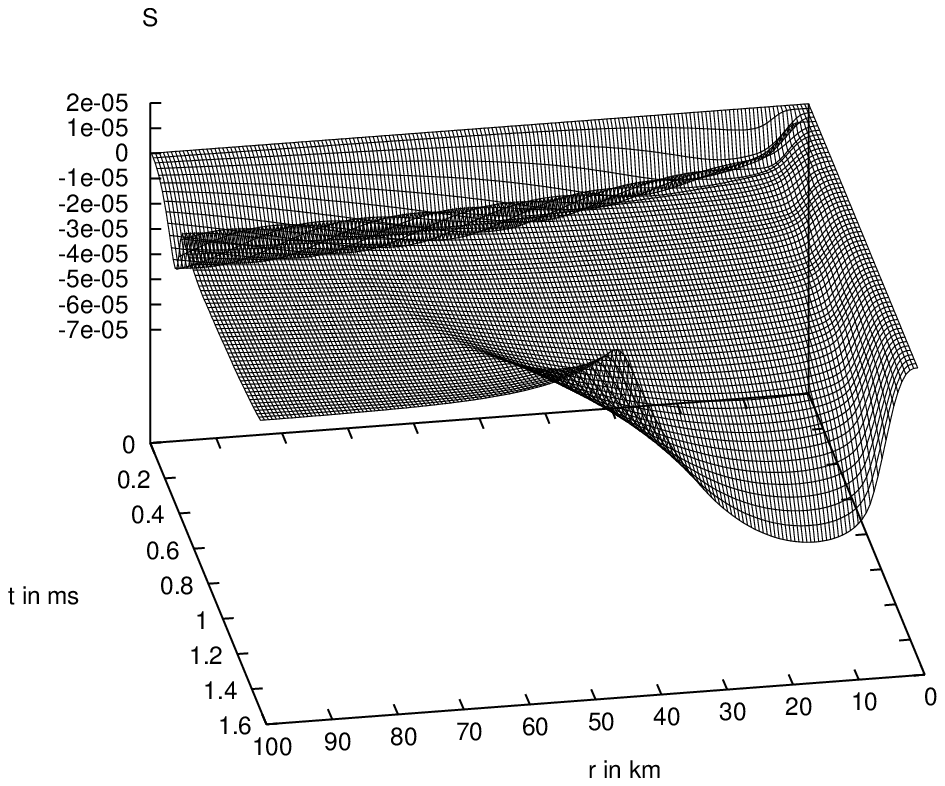}
\end{minipage}
\caption{\label{S3d}Evolution of $S$. The left panel shows the evolution
	for $T_0 = 0$, in the right panel $S_0 = 0$. In the first 
	case shows a burst of gravitational waves that propagate both in
	and outwards. The ingoing pulse gets reflected at the origin and
	travels back outwards again. In the other case, a wave originating
	from the interior of the neutron star travels outwards. Note
	the differences in the amplitudes of both cases.}
\end{figure}

\begin{figure}[hhh]
\leavevmode
\hspace*{-1cm}
\begin{minipage}{10cm}
\epsfxsize=\textwidth
\epsfbox{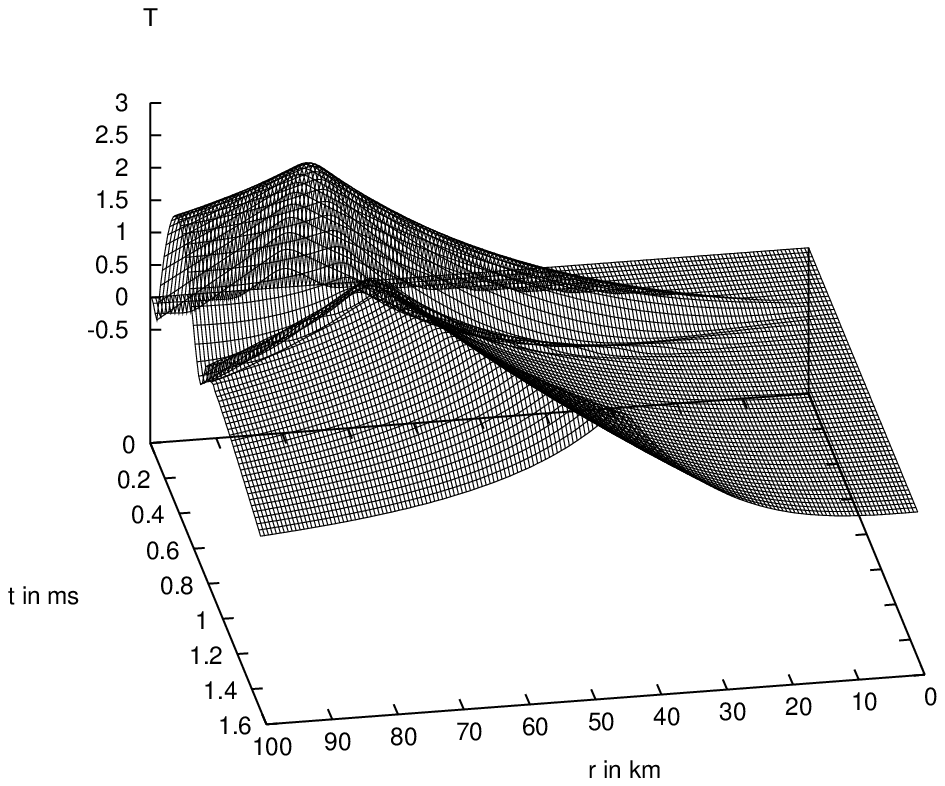}
\end{minipage}
\hspace*{-1.8cm}
\begin{minipage}{10cm}
\epsfxsize=\textwidth
\epsfbox{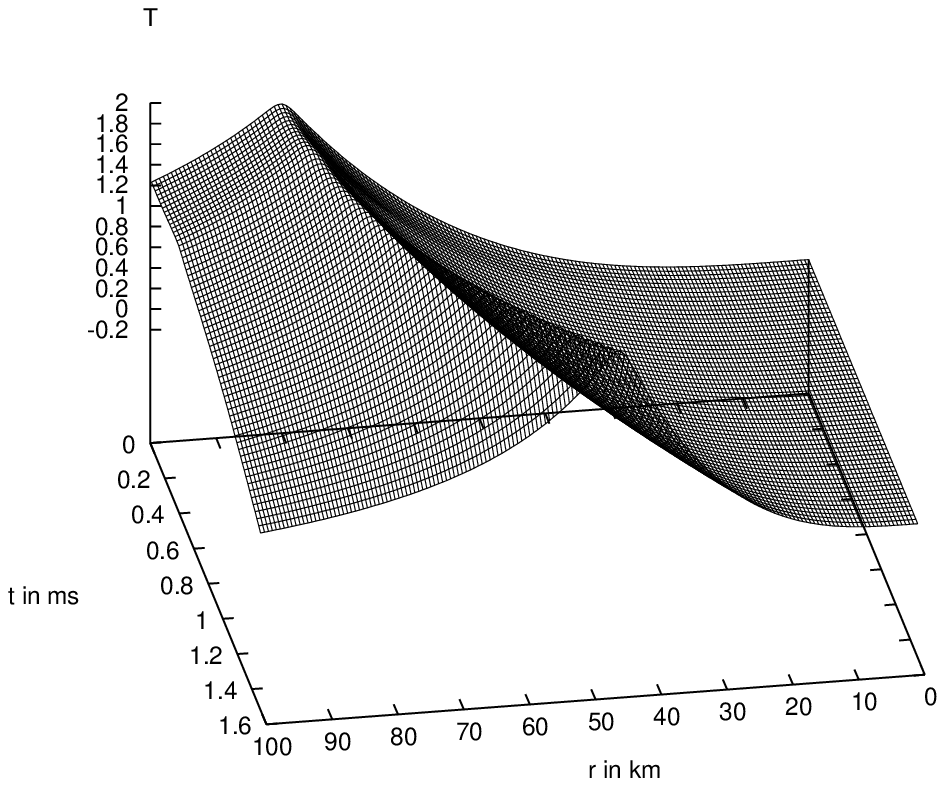}
\end{minipage}
\caption{\label{T3d} Evolution of $T$. The left panel shows the evolution
	for $T_0 = 0$, in the right panel $S_0 = 0$. In the first
	case, we can see the built up of $T$ which is disturbed by the 
	reflected part of the wave that was sent out by the particle.
	The second case clearly shows that $T$ basically has its ``right'' 
	shape right from the beginning.}
\end{figure}

\begin{figure}[hhh]
\leavevmode
\hspace*{-1cm}
\begin{minipage}{10cm}
\epsfxsize=\textwidth
\epsfbox{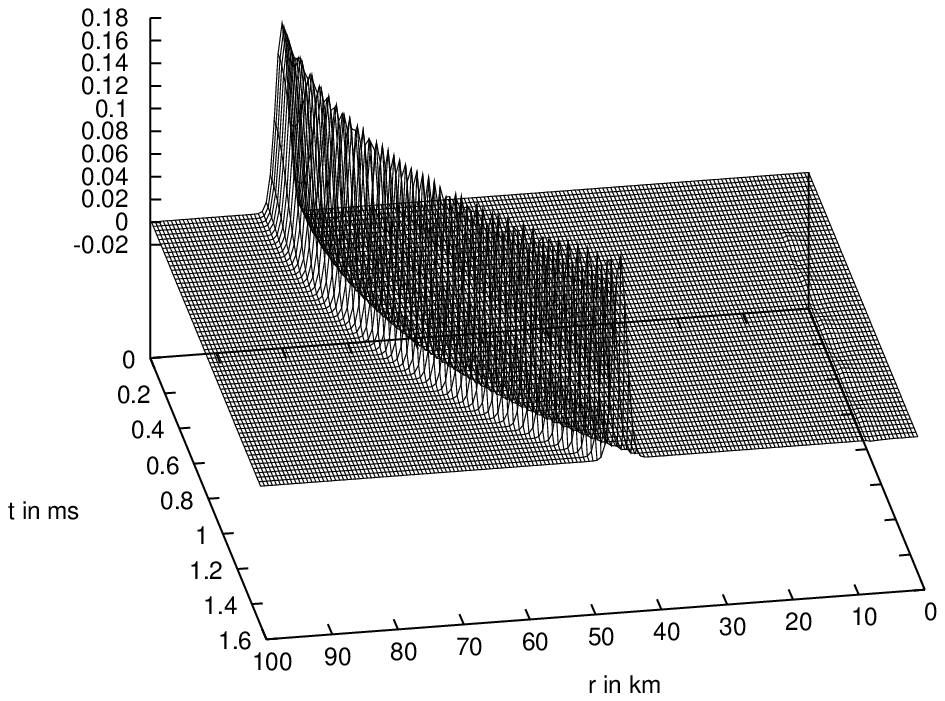}
\end{minipage}
\hspace*{-1.8cm}
\begin{minipage}{10cm}
\epsfxsize=\textwidth
\epsfbox{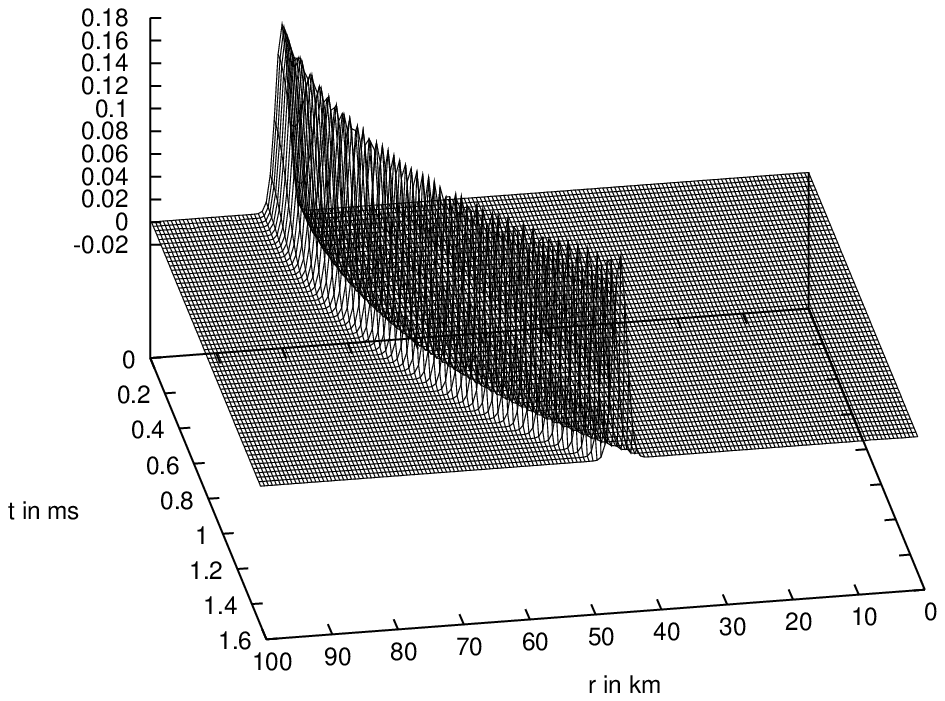}
\end{minipage}
\caption{\label{HC3d} Evaluation of the Hamiltonian constraint during the 
	evolution. Again, the left panel shows the evolution for 
	$T_0 = 0$, in the right panel $S_0 = 0$. The Gaussian shape of 
	the particle is	clearly visible and has been chosen to be very broad 
	for a better visualization. The particle is initially at rest
	and starts falling towards the neutron star.}
\end{figure}

\begin{figure}[hhh]
\leavevmode
\begin{minipage}{8.5cm}
\epsfxsize=\textwidth
\epsfbox{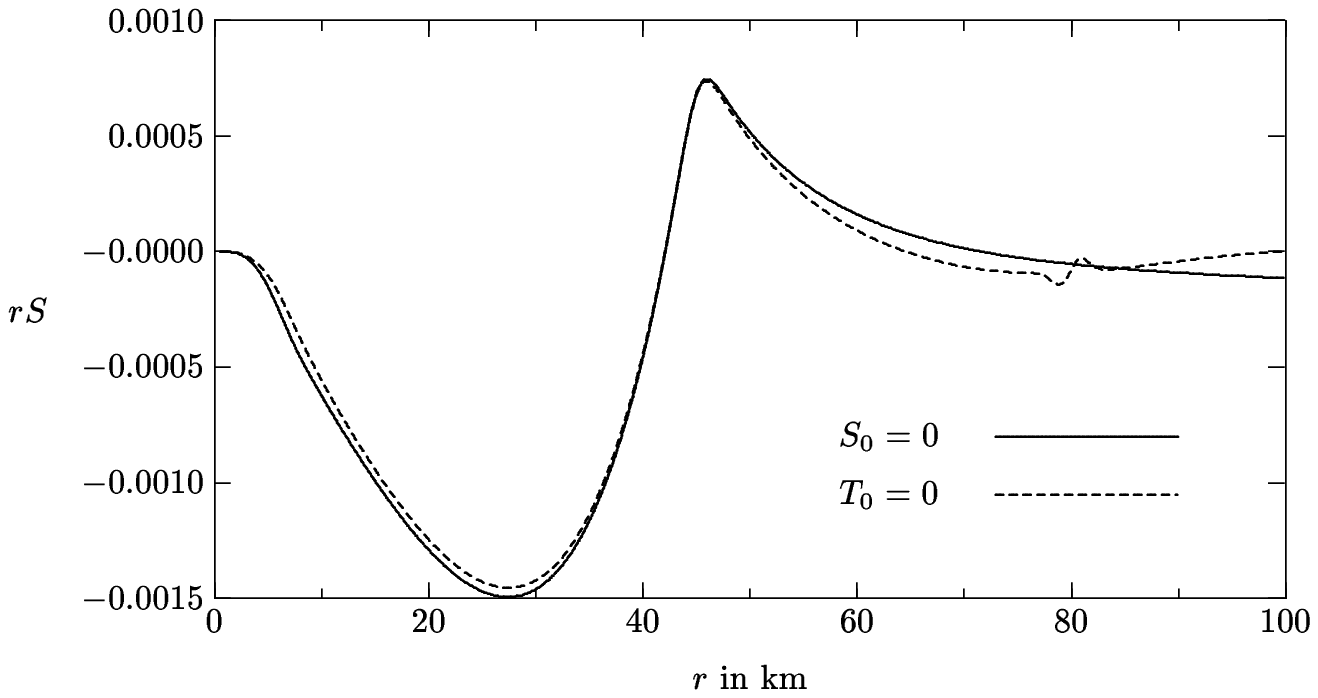}
\end{minipage}
\begin{minipage}{8.5cm}
\epsfxsize=\textwidth
\epsfbox{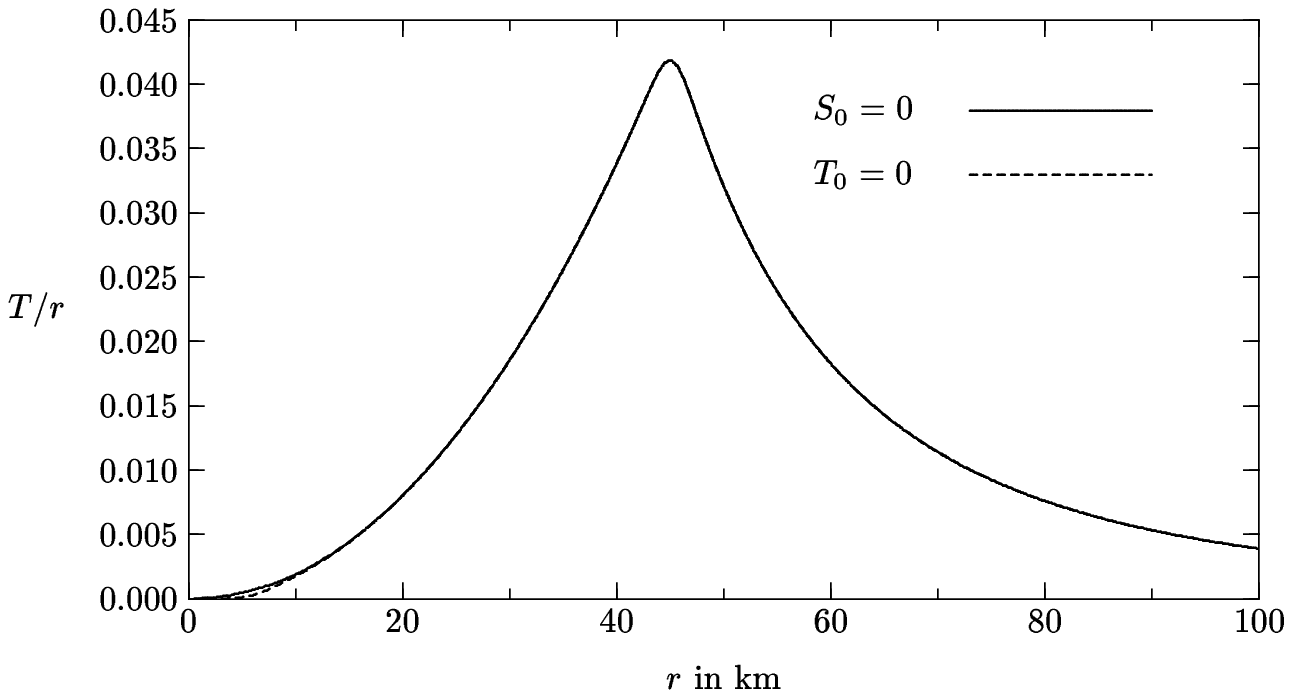}
\end{minipage}
\vspace*{3mm}
\caption{\label{STfinal} Plot of the variables $S$ and $T$ at the
end of the evolution of the different initial data. The artificial
radiation of the initial data has been radiated away and both
variables have assumed their ``right'' values that is independent of
the initial data. The difference in $S$ is due to the fact that the
initial data with $T_0 = 0$ contain much more radiation which
excites the neutron stars to pulsations which in turn disturb the
profile of $S$. The variable $T$ is mostly unaffected by this.}
\end{figure}

In Figs.~\ref{S3d} and \ref{T3d}, we show the evolution of the two
possible choices of initial data for a particle falling from rest. The
left panels show the evolution of $S$ and $T$ in the case $T_0=0$,
whereas the right panels show the evolution of $S$ and $T$ with
$S_0=0$.  The differences are obvious.  The initial shape of $T$ in
the latter case is almost unchanged during the evolution, whereas $S$
starts to acquire its ``right'' shape. In the other case, we can see a
huge burst of radiation propagating in both directions. In the same
time $T$ is acquiring its ``right'' shape. In Fig.~\ref{HC3d}, we show
the evolution of the Hamiltonian constraint where we evaluate the
left-hand side of (\ref{hc}) which monitors the ``path'' of the
particle. In both cases the graphs agree.  Lastly, in
Fig.~\ref{STfinal}, we also show the metric functions $S$ and $T$ at
the end of the evolution. This figure clearly show that, regardless of
the chosen initial data, the functions $S$ and $T$ will adjust to
their proper values after having radiated away the superfluous initial
wave content.

We should note that we cannot escape the whole ambiguity of how to
choose the variables $S$ and $T$ by using the Zerilli formalism,
instead. There, any regular initial data are valid initial data since
by construction the Zerilli function always satisfies the constraint
equations. Hence, we cannot see a priori whether or not the chosen
initial data will have additional radiation content.

If the particle is initially not at rest, in addition to solving the
Hamiltonian constraint (\ref{hc}), we also have to solve the momentum
constraints (\ref{mc1}) and (\ref{mc2}). Of course, as with the
Hamiltonian constraint, we are again faced with the same kind of
ambiguities in solving that equation.

Now, for a particle being initially at rest or very slow, or for
circular orbits, we do not worry about what kind of initial data we
choose. Since any initial gravitational wave pulse travels with the
speed of light and is therefore much faster than the particle, it will
long be gone when the particle comes close to the star. However, if
the particle's initial velocity is close to the speed of light, then
the particle ``rides'' on its own wave pulse, and it will be not clear
any more whether the excitation of some particular modes of the
neutron star is due the the particle itself or due to the initial
burst which comes from inappropriate initial data. If the particle
is slow enough, those two effects can clearly be distinguished. For
very fast particles, this is not possible any more.  This is
particularly bothersome given that a pulse of gravitational waves will
predominately excite $w$-modes. If we have a wave signal from a
particle that grazes a neutron star with almost the speed of light,
and we find that, indeed, there are some traces of a $w$-mode, it
would be very difficult to state that the excitation of $w$-modes is a
``real'' signal and not an artifact due to the inappropriate initial
data.

To obtain an approximate answer, we again turn to the flat space case.
In this limit, the equations governing the evolution of $S$ and $T$
reduce to two simple coupled wave equations with a source term for each
that takes into account the presence of the particle:
\begin{align}\label{Sw}
	\dff{S}{t} &= \dff{S}{r} - \frac{l(l+1)}{r^2}\,S
	+ 16\pi\,\mu E\,\frac{v^2}{r^3}\,\delta(r-R(t))Y^*_{lm}\\
	\dff{T}{t} &= \dff{T}{r} - \frac{l(l+1)}{r^2}\,T + 4\,S
	- 8\pi\,\frac{\mu}{rE}\,\delta(r-R(t))Y^*_{lm}\;.\label{Tw}
\end{align}
Herein, we have omitted the terms proportional to $L$, since we let the
particle move on a radial trajectory with constant velocity $v$, hence
it is $L = 0$ and
\begin{align}
	R(t) &= R_0 + v\,t\;,
\end{align}
with $R_0$ being the initial location of the particle. Furthermore,
the normalized energy $E$ of the particle is just given by the Lorentz
factor
\begin{align}
	E &= \frac{1}{\sqrt{1 - v^2}}\;.
\end{align}
It is interesting to note that the wave equation for $S$ is totally
decoupled from the one for $T$. However the solutions of (\ref{Sw})
and (\ref{Tw}) have to satisfy the flat space Hamiltonian constraint
which reads
\begin{align}\label{HCflat}
	\dff{T}{r} - \frac{l(l+1)}{r^2}\,T - r\,\df{S}{r}
	- \(2 + \half l(l+1)\)S
	&= - 8\pi\,\frac{\mu E}{r}\,\delta(r-R(t))Y^*_{lm}\;.
\end{align}
We now seek an exact solution of (\ref{Sw}) that obeys the right 
boundary conditions at the origin and at infinity. Once found, we may
use (\ref{HCflat}) to numerically compute the appropriate $T$. We state
that a solution for (\ref{Sw}) is given by the following series ansatz:
\begin{align}\label{Sansatz}
	S(t,r) &= A\(\sum_{i = 0}a_i\,
	\frac{r^{2i+l+1}}{(R_0 + v\,t)^{2i+l+3}}\Theta(R_0 + v\,t - r)
	+ \sum_{i = 0}b_i\,\frac{(R_0 + v\,t)^{l-2i-2}}{r^{l-2i}}
	\Theta(r - R_0 - v\,t)\)\;,
\end{align}
where $\Theta$ is the Heaviside function which satisfies
\begin{align}
	\Theta(x) &= \begin{cases}
	0 ,& x < 0\\
	1 ,& x \ge 0\\
	\end{cases}\;.
\end{align}
Continuity at $r = R_0 + v\,t$ requires that
\begin{align}\label{ab}
	\sum_{i = 0}a_i &= \sum_{i = 0}b_i\;.
\end{align}
The overall amplitude will be determined by $A$, hence we deliberately 
may set
\begin{align}
	\sum_{i = 0}a_i &= \sum_{i = 0}b_i \,=\, 1\;.
\end{align}
The amplitude $A$ and the coefficients $a_i$ and $b_i$ can be found
by plugging (\ref{Sansatz}) into (\ref{Sw}). For $A$ we find
\begin{align}\label{S0}
	A &= \frac{16\pi\,\mu E^3 v^2\,Y^*_{lm}}{2\,l + 1
	+ 2\sum_{i = 0}\,i\(a_i - b_i\)}\;,
\end{align}
and the coefficients $a_i$ and $b_i$ are determined by the following
recursion relations
\begin{align}
	a_{i+1} &= a_i\,v^2\,\frac{\(2\,i + l + 3\)\(2\,i + l + 4\)}
	{2\(i + 1\)\(2\,i + 2\,l + 3\)}\\
	b_{i+1} &= b_i\,v^2\frac{\(l - 2\,i - 2\)\(l - 2\,i - 3\)}
	{2\(i + 1\)\(2\,i - 2\,l + 1\)}\;.
\end{align}
It is interesting to note that, while the series in $a_i$ never
terminates, the series in $b_i$ always terminates because one of the
two factors in the numerator will become zero for some $i$. The series
in $a_i$ converges if and only if $|v| < 1$.

\begin{figure}[hhh]
\leavevmode
\hspace*{-3mm}
\begin{minipage}{8.7cm}
\epsfxsize=\textwidth
\epsfbox{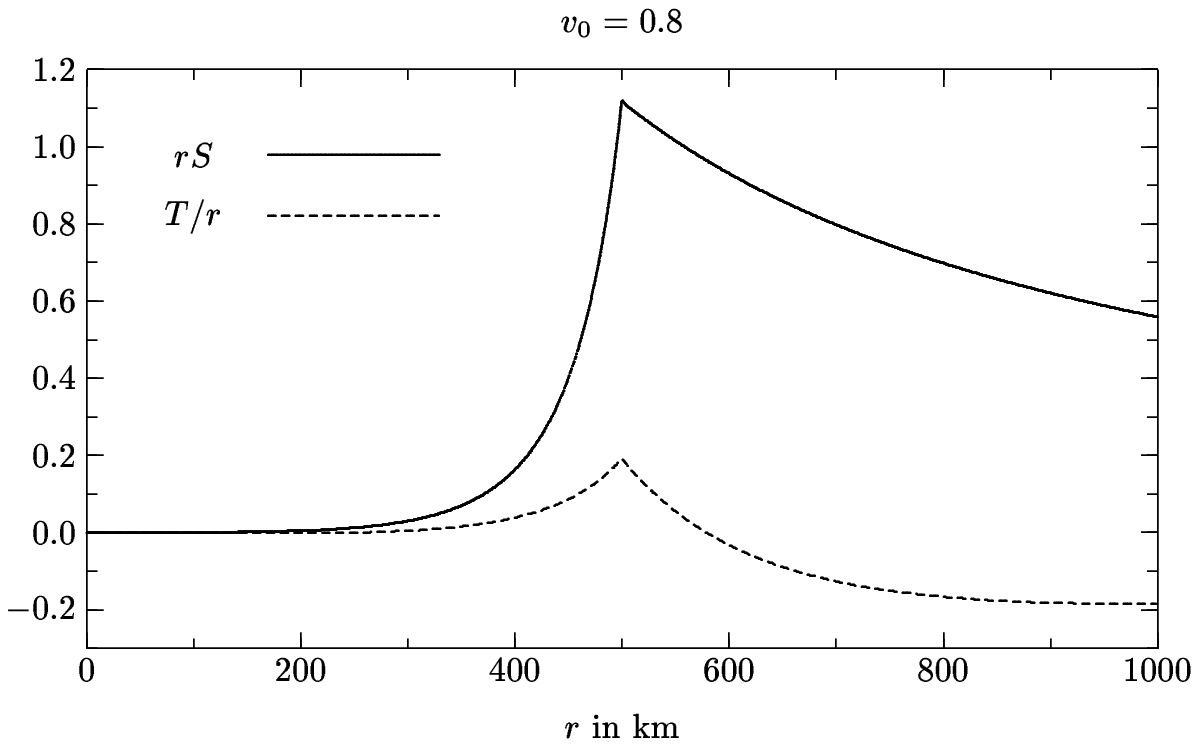}
\end{minipage}
\hspace*{2mm}
\begin{minipage}{8.7cm}
\epsfxsize=\textwidth
\epsfbox{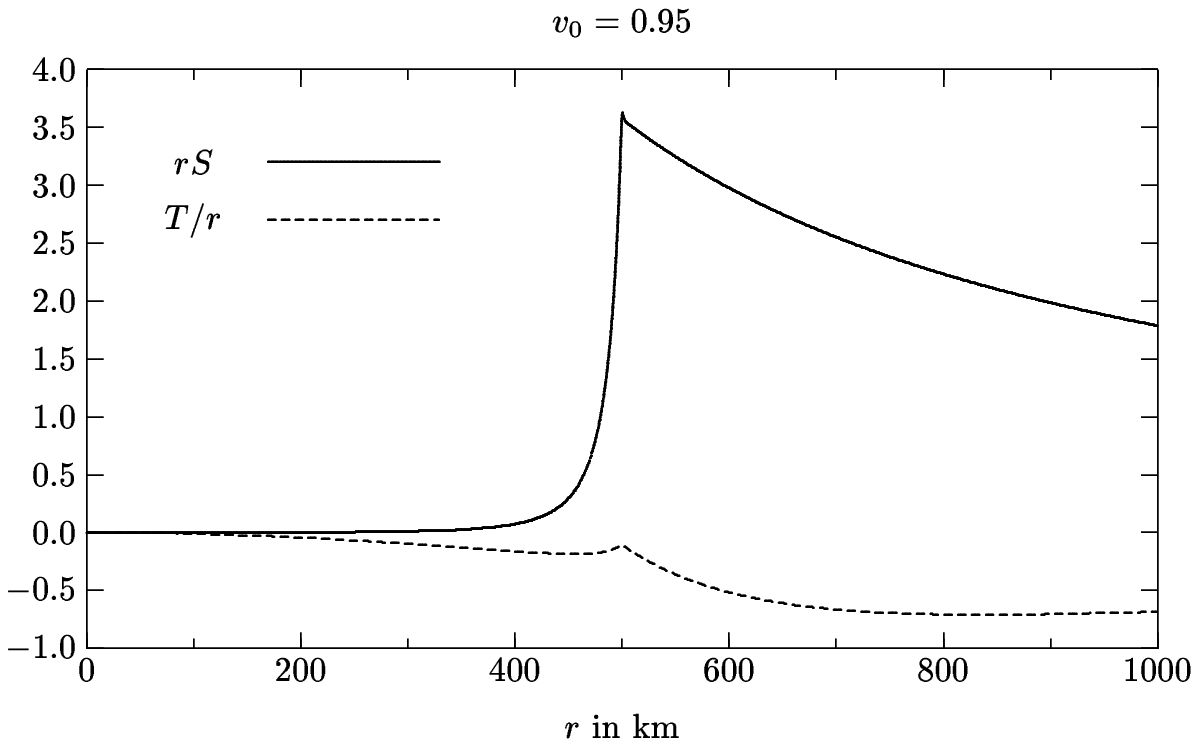}
\end{minipage}
\vspace*{3mm}
\caption{\label{analytic}Initial data for a particle with different initial
radial velocities $v$ where $S$ is given by (\ref{Sansatz}) and $T$ is
obtained by solving (\ref{HCflat}). For small velocities, $T/r$
dominates $rS$ whereas for ultra-relativistic velocities, $rS$
dominates $T/r$.}
\end{figure}

In Fig.~\ref{analytic}, we show the initial data obtained from
(\ref{Sansatz}) for a particle that is located at $R_0 = 500\,$km with
different initial velocities. Here, instead of $S$ and $T$, we plot
the quantities $rS$ and $T/r$, since from (\ref{hij_polar}) it is
clear that only those expressions allow direct comparisons. For $v=0$
it is $rS=0$ but the amplitude of $rS$ grows rapidly when the
particle's velocity approaches the speed of light, whereas the peak of
$T/r$ slightly decreases. In the ultra-relativistic limit $rS$ totally
dominates over $T/r$.

The solutions above for $S$ and $T$ are not valid any more
if we consider the particle in curved spacetime because they would
violate the constraints. However, we can still use (\ref{Sansatz}) as
a prescription for $S$ and then use the curved space Hamiltonian
constraint (\ref{hc}) to solve for $T$. Furthermore, we can compute
$K$ from $K = dS/dt$ and then use the momentum constraint
(\ref{MC2nd}) to compute $K_5$. As long as the particle does not have
any angular momentum and is far away from the neutron star, the thus
obtained initial values should be a good approximation for a boosted
particle on a Schwarzschild background. But if the particle has a
large angular momentum, there will be additional source terms in the
evolution equations and our approximation should break down. However,
we are mainly interested in trajectories which come very close to the
neutron star, and, hence, have only small angular momentum. In this
case the above prescription still yields good initial data.

\section{Numerical results}
\label{sec:res}

\begin{figure}[hhh]
\hspace*{3cm}
\begin{minipage}{10cm}
\leavevmode
\epsfxsize=\textwidth
\epsfbox{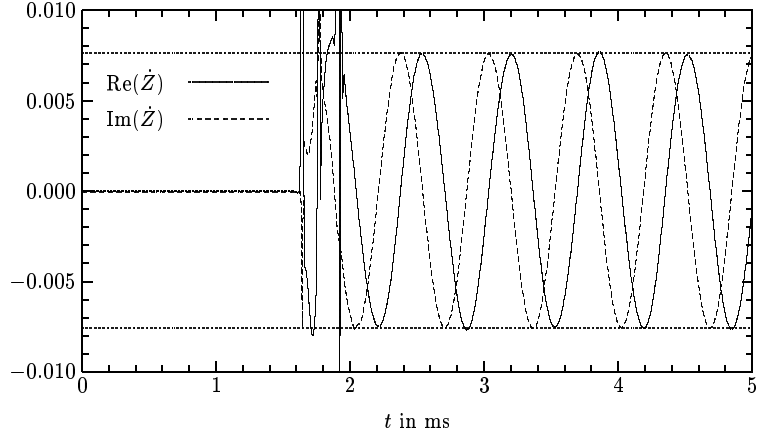}
\vspace*{3mm}
\caption{\label{circ}Evolution of the real and imaginary parts of
$\dot{Z}_{22}$ at $r = 500$km. After the initial radiation
bursts the waveforms show periodical oscillations with twice the
orbital frequency. The amplitude is about $0.0076$ which corresponds
to a radiation power of $(M/\mu)^2\,dE_{22}/dt = 5.46\times
10^{-5}$.}
\end{minipage}
\end{figure}

To compare the code with known results, we first consider a particle in a
circular orbit around the neutron star and compute the radiated energy
at infinity. Numerically, this can be accomplished by evaluating the
Zerilli function at some large distance. In terms of the metric
variables $S$ and $T$, the Zerilli function reads
\begin{align}\label{zerilli}
	Z &= - \frac{2}{l(l+1)}
	\frac{1 - \frac{2M}{r}}{l(l+1) - 2 + \frac{6M}{r}}
	\(2\,rT' + \frac{2M - r\(2 + l(l+1)\)}{r - 2M}\,T - 2\,r^2S\)\;.
\end{align}
The radiated power for some particular values of $l$ and $m$ can then be
computed from
\begin{align}
	\frac{dE_{lm}}{dt} &= \frac{1}{64\pi}\frac{(l+2)!}{(l-2)!}\,
	|\frac{Z_{lm}}{dt}|^2\;.
\end{align}
Cutler et al.~\cite{CFPS93} have numerically computed the normalized
gravitational power $(M/\mu)^2\,dE_{lm}/dt$ radiated by a particle
orbiting a Schwarzschild black hole for various values of $l, m$ and
$R_0/M$. In table II, they show the multipole components for $R_0/M =
10$. To compare the output of our code to those results, we chose the
mass of the neutron star to be $M = 1.99$\,km and the orbit of the
particle the be at $R_0 = 19.9\,$km in order to obtain the ratio of
$R_0/M = 10$. The mass of the particle was set to $\mu = 1\,$km. In
Fig.~\ref{circ}, we show $dZ_{lm}/dt$ as a function of time extracted
at $r = 500\,$km for $l=m=2$. After some wave bursts that come from
the inappropriate choice of initial data, we see that the signal is
periodic with a frequency of twice the orbital frequency of the
particle. The amplitude is about $0.0076$ which corresponds to a
radiated power of $(M/\mu)^2\,dE_{22}/dt = 5.46\times 10^{-5}$ which
is in excellent agreement with Cutler et al. who obtain a value of
$(M/\mu)^2\,dE_{22}/dt = 5.388\times10^{-5}$. The slightly higher
value of our result may be due to the fact that at $r = 500\,$km the
Zerilli function is still somewhat off its asymptotic value which will
be a little bit smaller. And, of course, a neutron star is not a black
hole, since its oscillation spectrum is quite different. But as long
as the particle does not excite the eigenmodes in a significant way,
i.e.~as long as the particle's orbital frequency is not in resonance
with one of the eigenmodes, the emitted radiation should be quite
similar to the black hole case. And, indeed, we find that we agree
with all the even and odd parity modes compiled in table II of Cutler
et al.~within a few percent, only for the case $l=3$ and $m=1$ we find
the radiated power to be $5.9\times10^{-10}$ instead of their value of
$5.71\times10^{-8}$.

If the orbital frequency is in resonance with one of the
eigenmodes, the emitted radiation can drastically increase
\cite{Koj87}. It is clear, however, that only the low-frequency modes
such as the $f$-mode can be excited by this mechanism, the frequencies of
the $w$-modes are much too high to be considerably excited. Besides,
due to their strong damping, there is no sharp resonance for the
$w$-modes at all, it is only for the weakly damped fluid modes that
resonant mode excitation can occur.

Therefore, the only way to possibly excite $w$-modes are very
eccentric orbits, where the periastron is very close to the surface of
the star or scattering processes with very small impact parameters.
The investigation of the latter is the main objective of this work.
Before presenting the results, we should point out some
problems that are related with the use of an evolution code.

\begin{figure}[hhh]
\leavevmode
\begin{minipage}{8.5cm}
\epsfxsize=\textwidth
\epsfbox{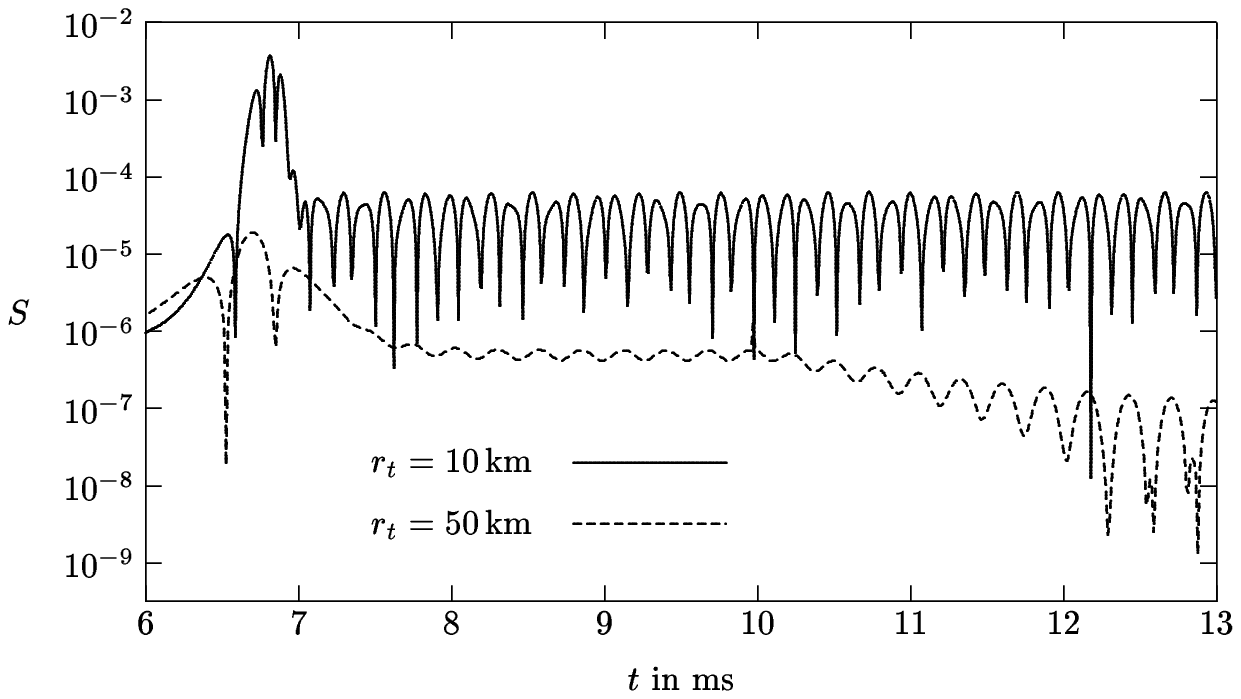}
\end{minipage}
\hspace*{2mm}
\begin{minipage}{8.5cm}
\epsfxsize=\textwidth
\epsfbox{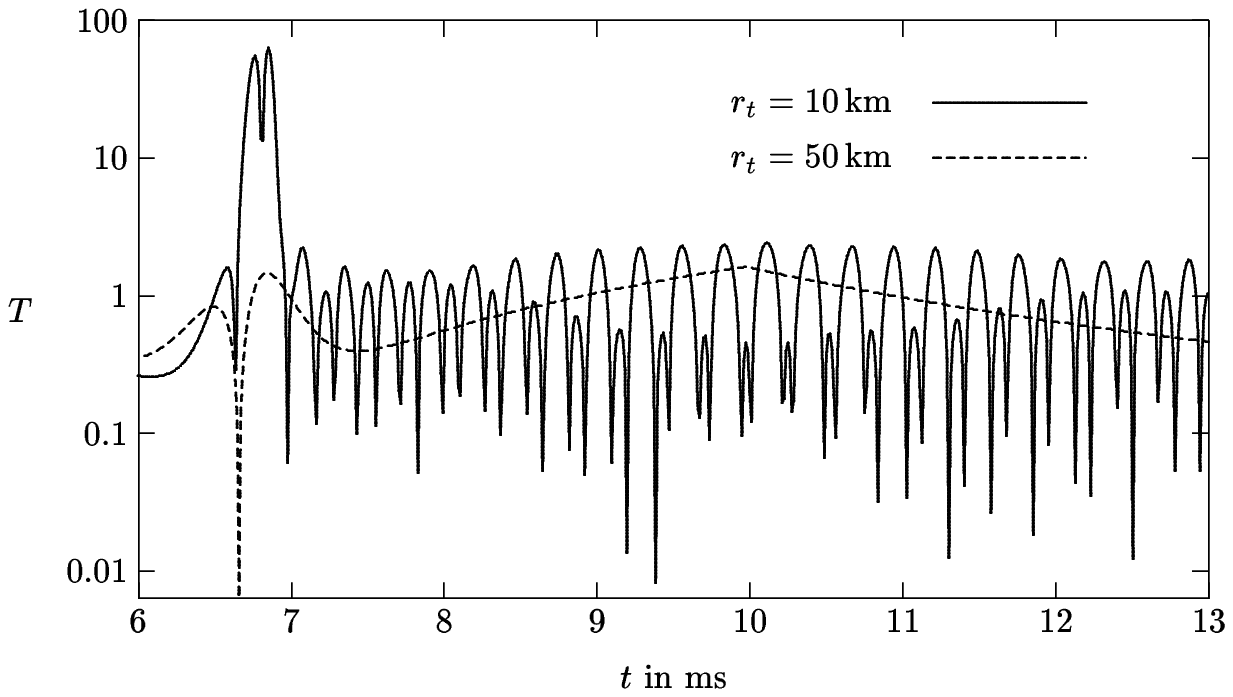}
\end{minipage}
\vspace*{3mm}
\caption{\label{fig:ST_v0=0.5b}shows the difference of excitation strengths
for the two different turning radii $r_t = 10$km (solid line) and $r_t
= 50$km (dashed line). The waveform of $S$ (left panel) for $r_t =
10$km is totally unaffected by the presence of the particle whereas
for $r_t = 50$km, the gravitational field shifts the amplitude of the
signal to higher values. For $T$ this effect is much more pronounced
and can already be detected for $r_t = 10$km. For $r_t = 50$km, the
oscillations of the neutron star are totally buried in the gravitational
field of the particle. At $t = 10$ms the particle crosses the observer
who is located at $r_{obs} = 1000$km.}
\end{figure}

Ideally, one should extract the waveform at $r = \infty$ which is
clearly not possible on a finite grid. We have to record the waveform
at some finite distance from the star. But this means that if the
particle moves on an unbounded orbit it will eventually cross the
location of the observer who will be located on a finite radius
$r_{obs}$. Of course, the particle is always slower than the
propagation speed of the gravitational waves, hence the observer will
see the wave signal first before the particle crosses
$r_{obs}$. However, when the particle is very fast and the observer is
not far enough away from the neutron star, then the signal and the
particle will cross the observer almost at the same time. That means
that the signal will be a superposition of the ``real'' gravitational
wave signal and the influence of the gravitational field of the
particle itself. The farther the observer moves outwards the better
the separation of the two components in the signal can be
made. Furthermore, the amplitude of the signal remains constant for
increasing $r$, whereas the influence of the particle decreases at
least as $1/r$. The effectiveness of this undesired influence of the
particle strongly depends on the actual excitation strength of the
neutron star oscillations. Smaller turning points $r_t$ of the
particle will induce oscillations with higher amplitudes and thus make
the influence of the particle undetectable. For large $r_t$ the
induced fluid oscillations are so weak that they will totally be
buried within the gravitational field of the particle. Figure
\ref{fig:ST_v0=0.5b} drastically shows the influence of the presence
of the particle on the observed waveform. Here, we compare the signals
that result for two scattering orbits with $r_t = 10\,$km and $r_t =
50\,$km. One can also see that the effect is much more pronounced in
the waveform of $T$. But regardless of this unwanted feature it is
still possible to infer qualitative statements from the obtained
wave signals.

Unfortunately, the influence of the source terms is even much worse for
the Zerilli function, which should be used to present the results,
since it is a gauge invariant quantity and directly related to the
radiated power. This is not the case for our variables $S$ and
$T$. But since we are mainly interested in assessing which modes are
excited by the scattering process, we do not have to compute the
Zerilli function. It is enough to just examine the waveforms of $S$
and $T$.

\begin{figure}[hhh]
\leavevmode
\begin{minipage}{8.6cm}
\epsfxsize=\textwidth
\epsfbox{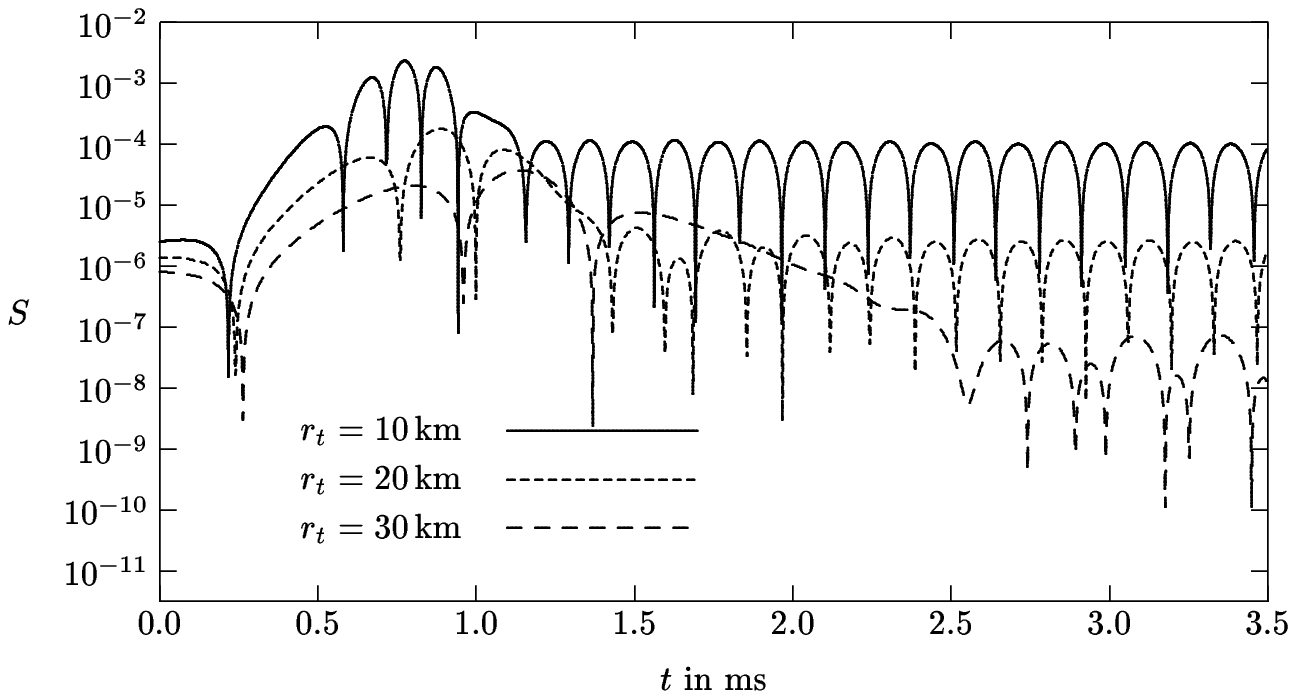}
\end{minipage}
\hspace*{2mm}
\begin{minipage}{8.5cm}
\epsfxsize=\textwidth
\epsfbox{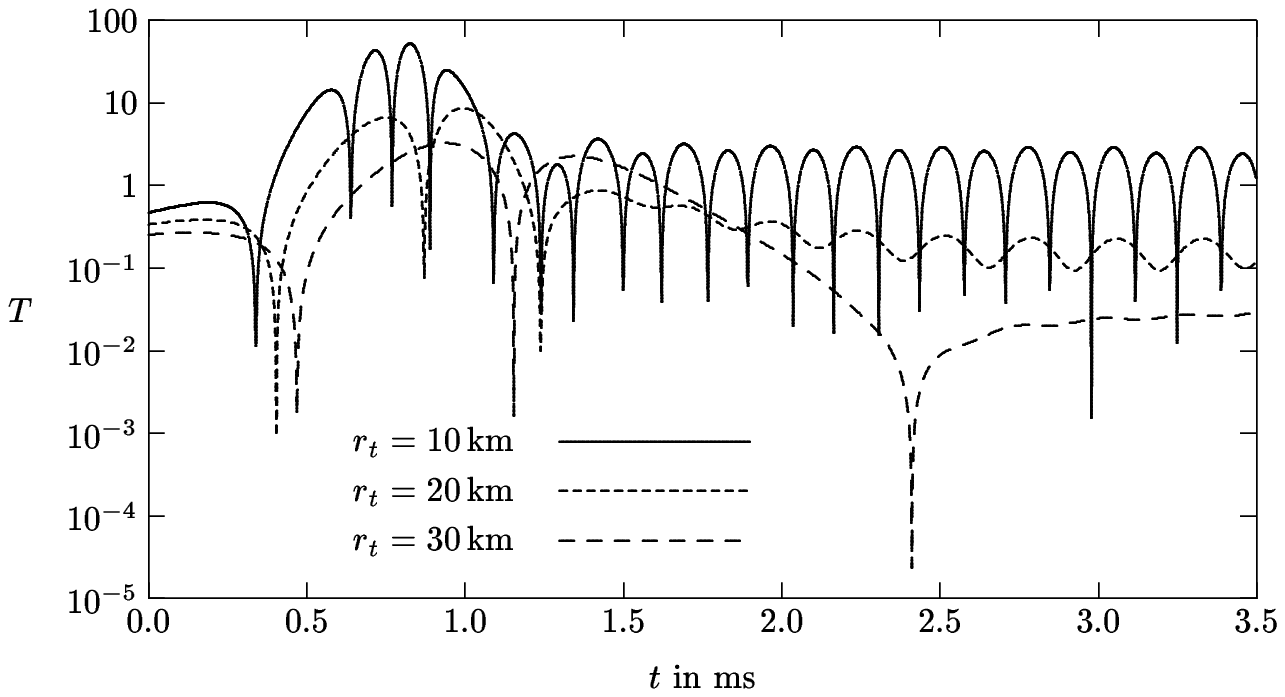}
\end{minipage}
\vspace*{3mm}
\caption{\label{fig:ST_v0=0.1}Waveforms of $S$ and $T$
for the three different turning radii $r_t = 10$km, $r_t = 20$km and
$r_t = 30$km. The initial velocity of the particle is $v_0 = -0.1$}
\end{figure}

\begin{figure}[hhh]
\leavevmode
\begin{minipage}{8.6cm}
\epsfxsize=\textwidth
\epsfbox{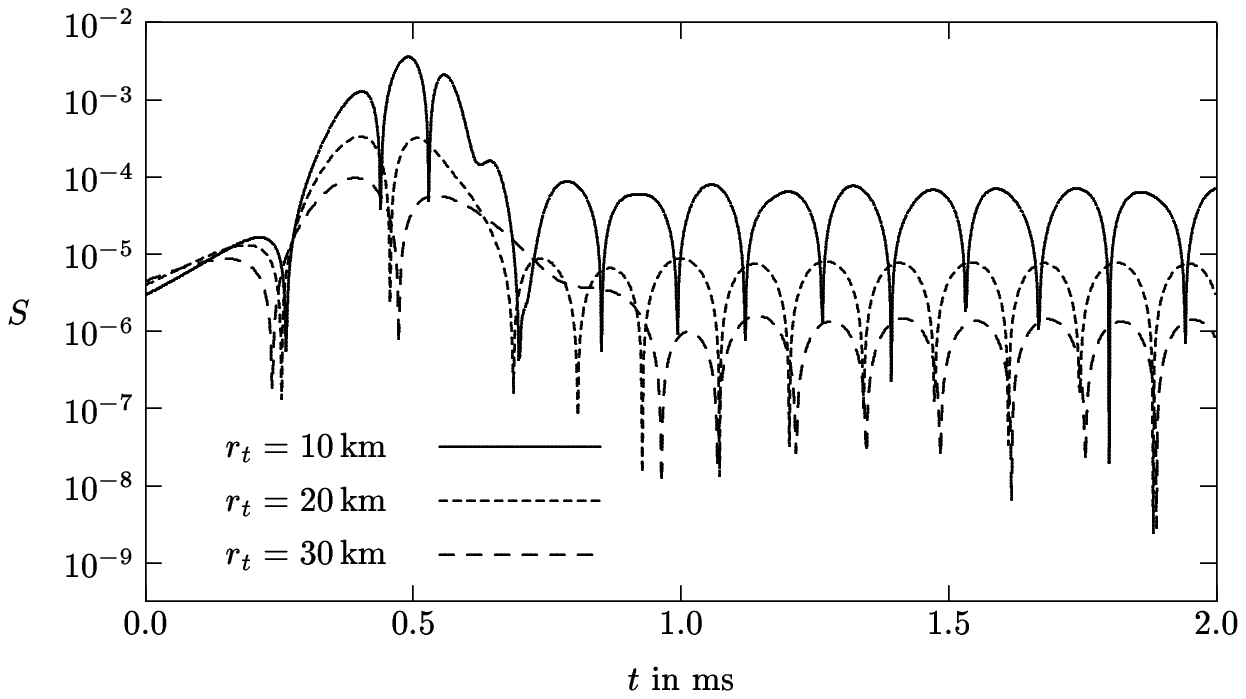}
\end{minipage}
\hspace*{2mm}
\begin{minipage}{8.5cm}
\vspace*{1mm}
\epsfxsize=\textwidth
\epsfbox{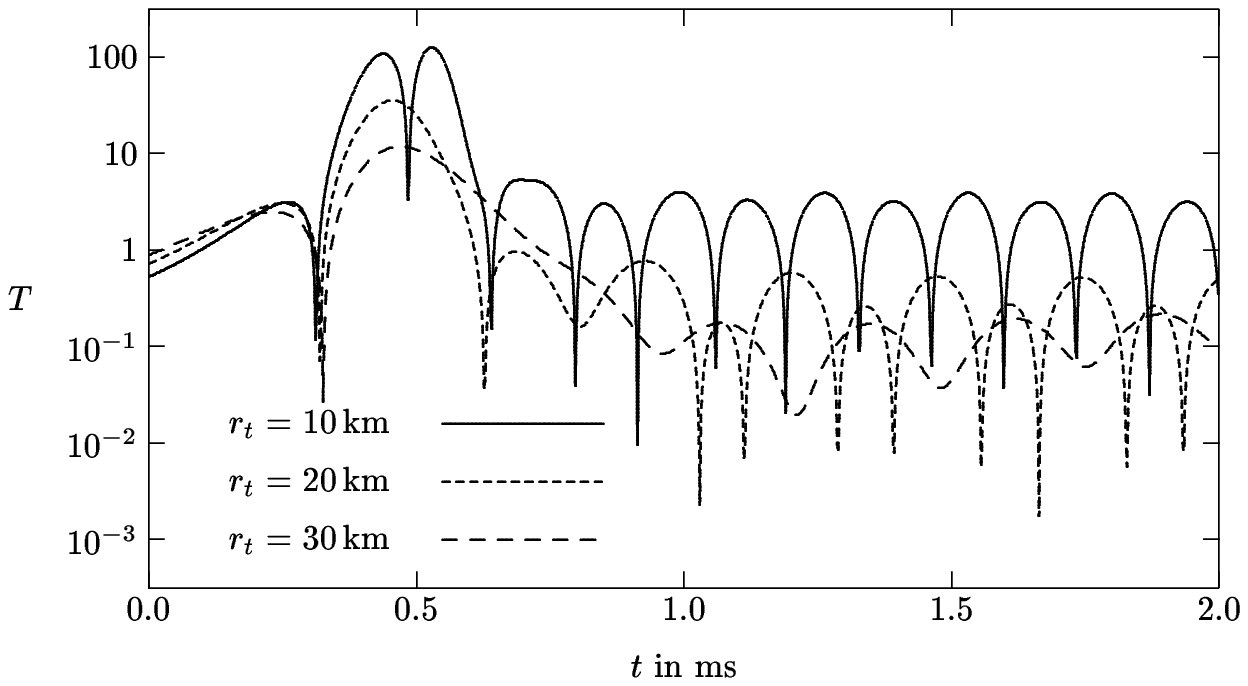}
\end{minipage}
\vspace*{3mm}
\caption{\label{fig:ST_v0=0.5a}Waveforms of $S$ and $T$
for the three different turning radii $r_t = 10$km, $r_t = 20$km and
$r_t = 30$km. The initial velocity of the particle is $v_0 = -0.5$}
\end{figure}

\begin{figure}[hhh]
\leavevmode
\begin{minipage}{8.5cm}
\epsfxsize=\textwidth
\epsfbox{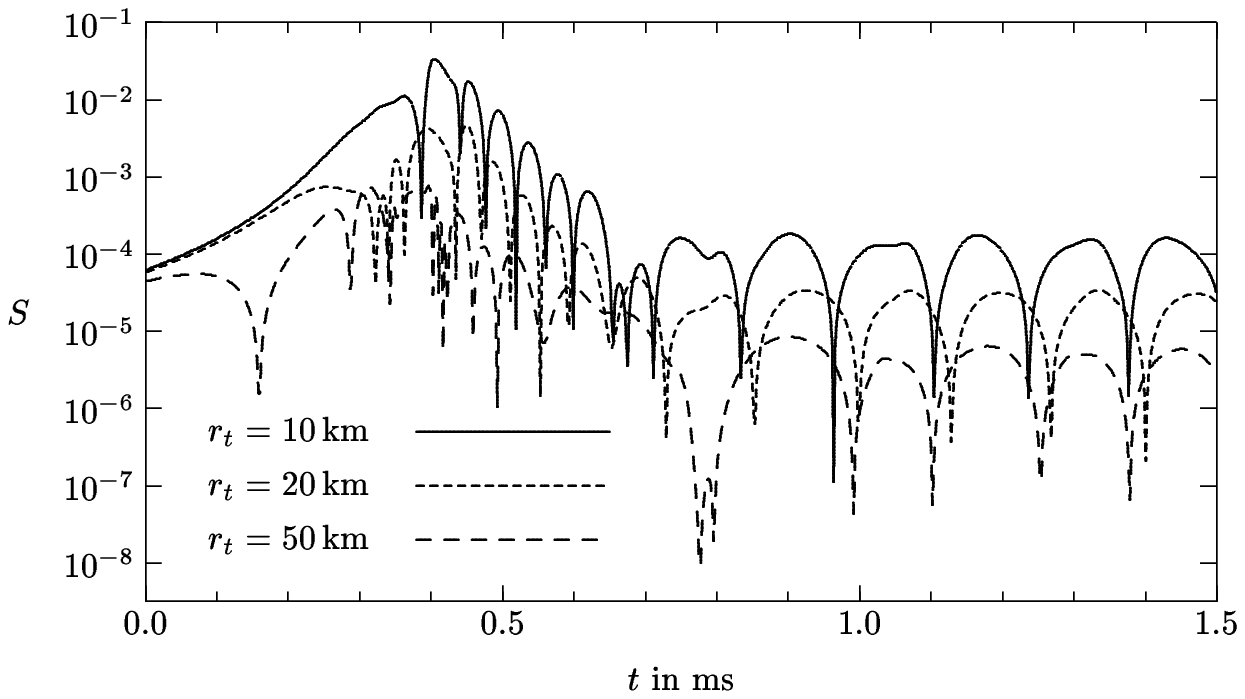}
\end{minipage}
\hspace*{3mm}
\begin{minipage}{8.5cm}
\vspace*{1mm}
\epsfxsize=\textwidth
\epsfbox{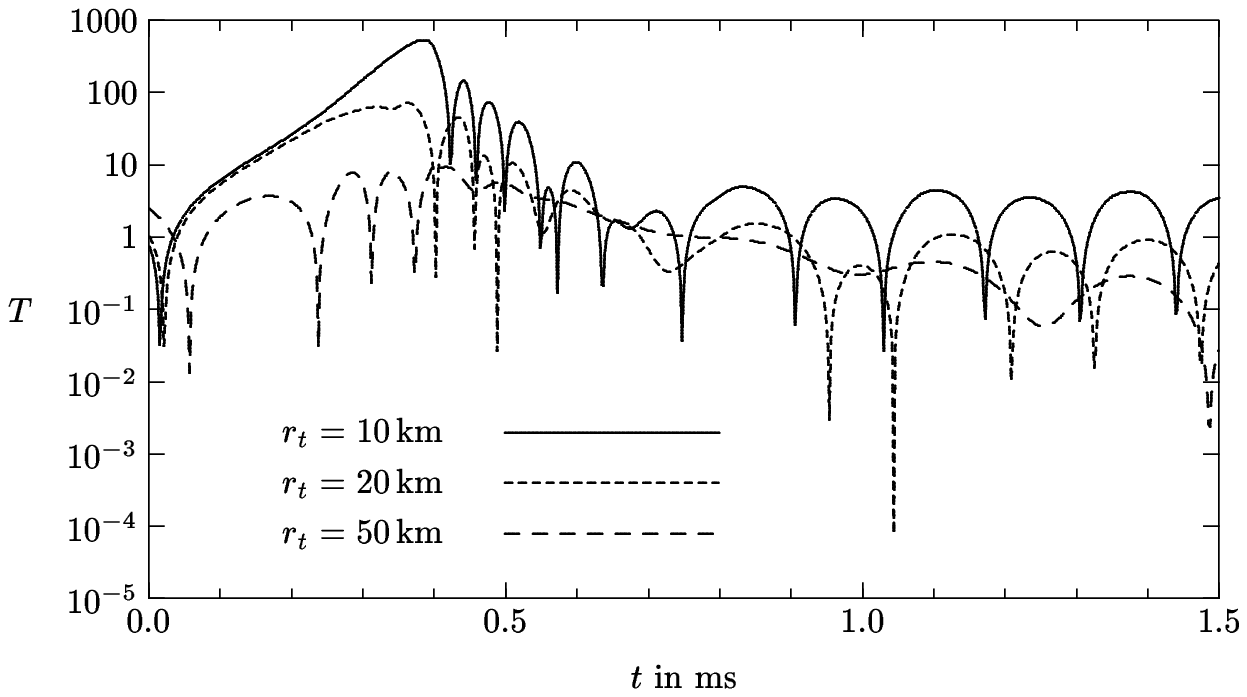}
\end{minipage}
\vspace*{3mm}
\caption{\label{fig:ST_v0=0.97}Waveforms of $S$ and $T$
for the three different turning radii $r_t = 10$km, $r_t = 20$km and
$r_t = 50$km. The initial velocity of the particle is $v_0 = -0.97$}
\end{figure}

\begin{figure}[hhh]
\leavevmode
\begin{minipage}{8.7cm}
\epsfxsize=\textwidth
\epsfbox{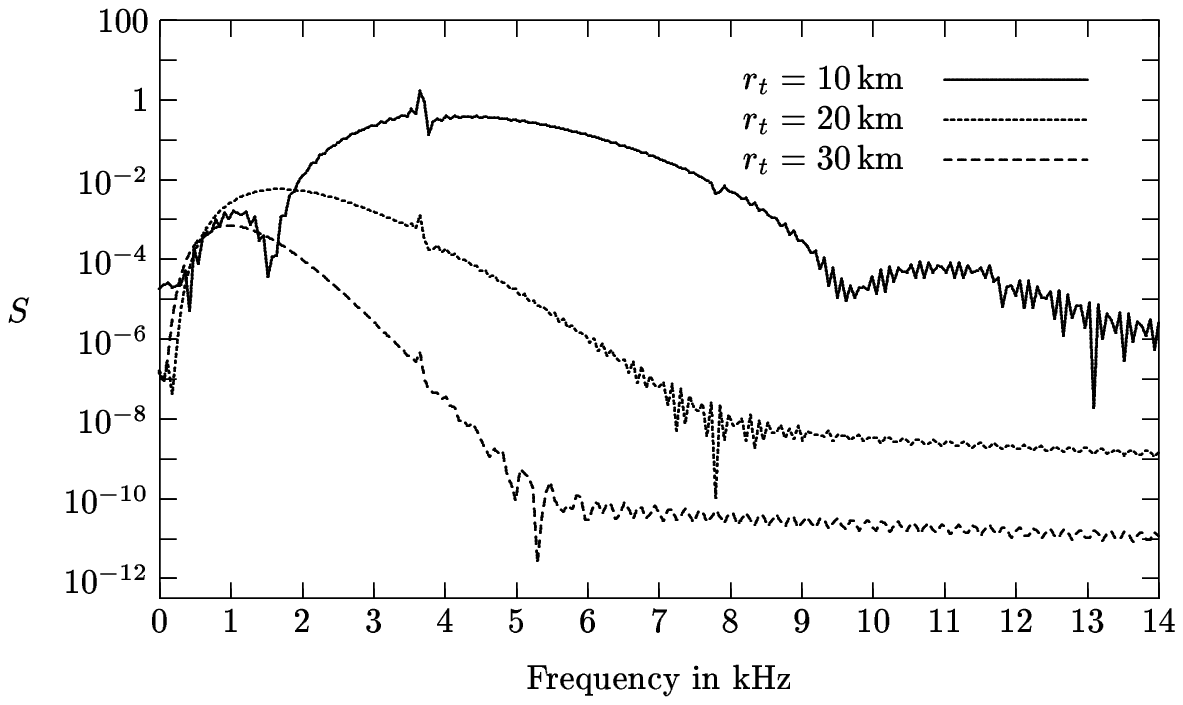}
\end{minipage}
\hspace*{2mm}
\begin{minipage}{8.5cm}
\epsfxsize=\textwidth
\epsfbox{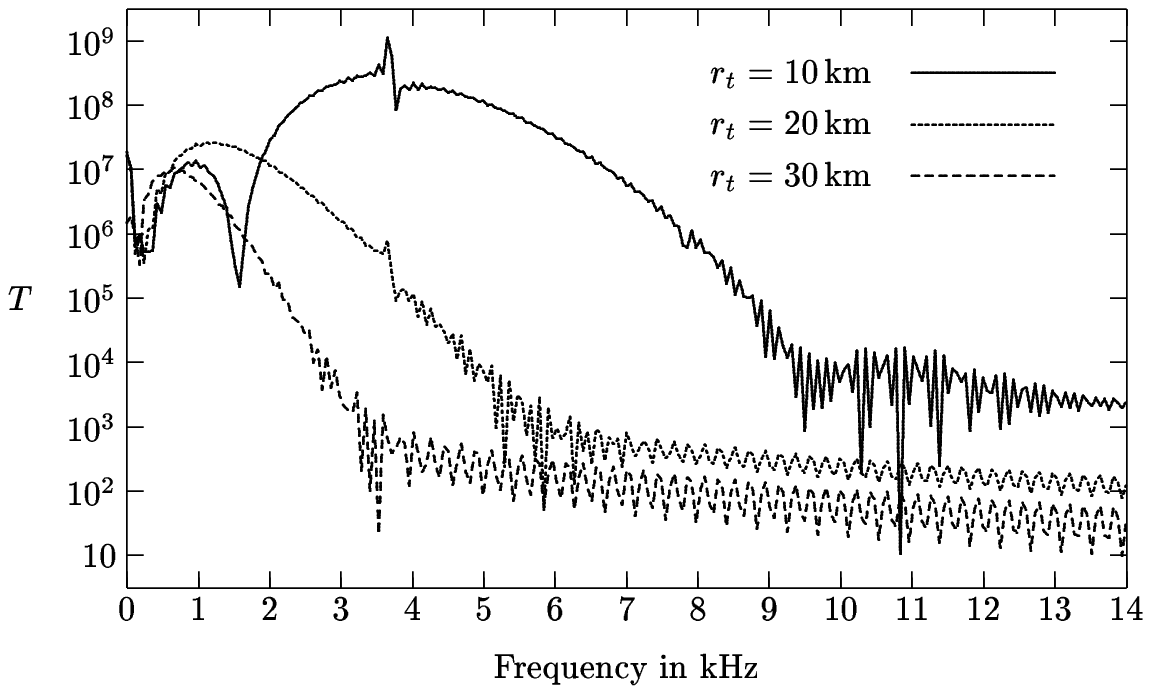}
\end{minipage}
\vspace*{3mm}
\caption{\label{fig:ST_v0=0.1fft}Fourier transform of the waveforms for
$v_0 = -0.1$}
\end{figure}

\begin{figure}[hhh]
\leavevmode
\begin{minipage}{8.7cm}
\epsfxsize=\textwidth
\epsfbox{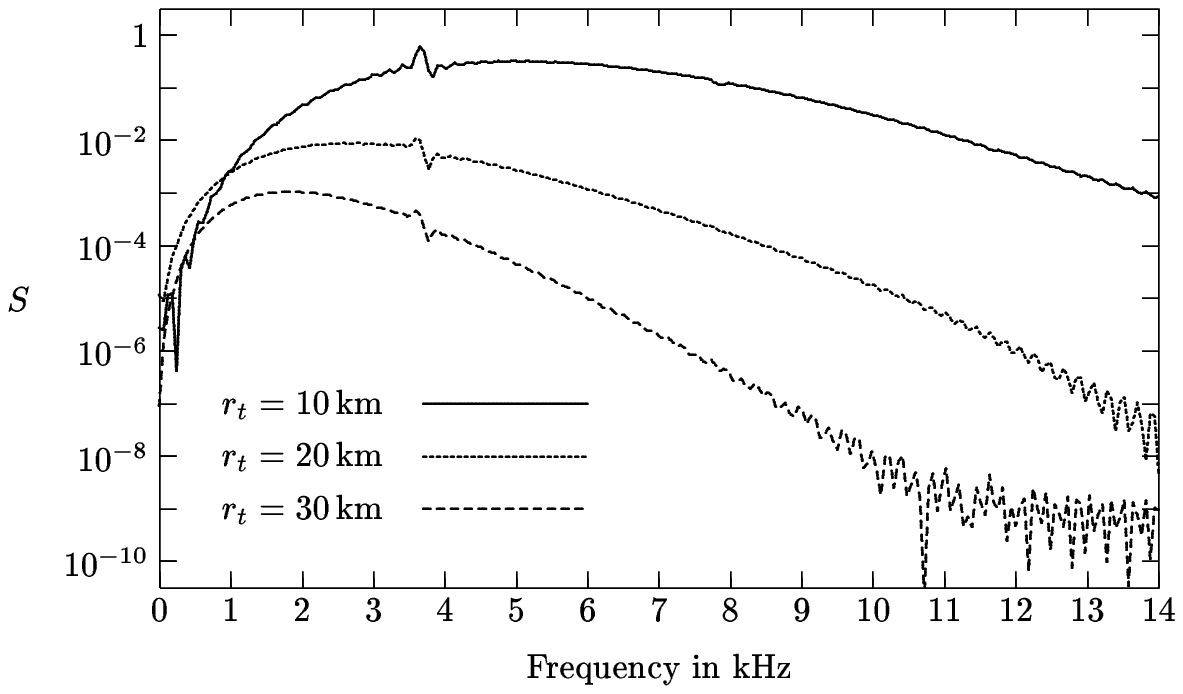}
\end{minipage}
\hspace*{2mm}
\begin{minipage}{8.5cm}
\epsfxsize=\textwidth
\epsfbox{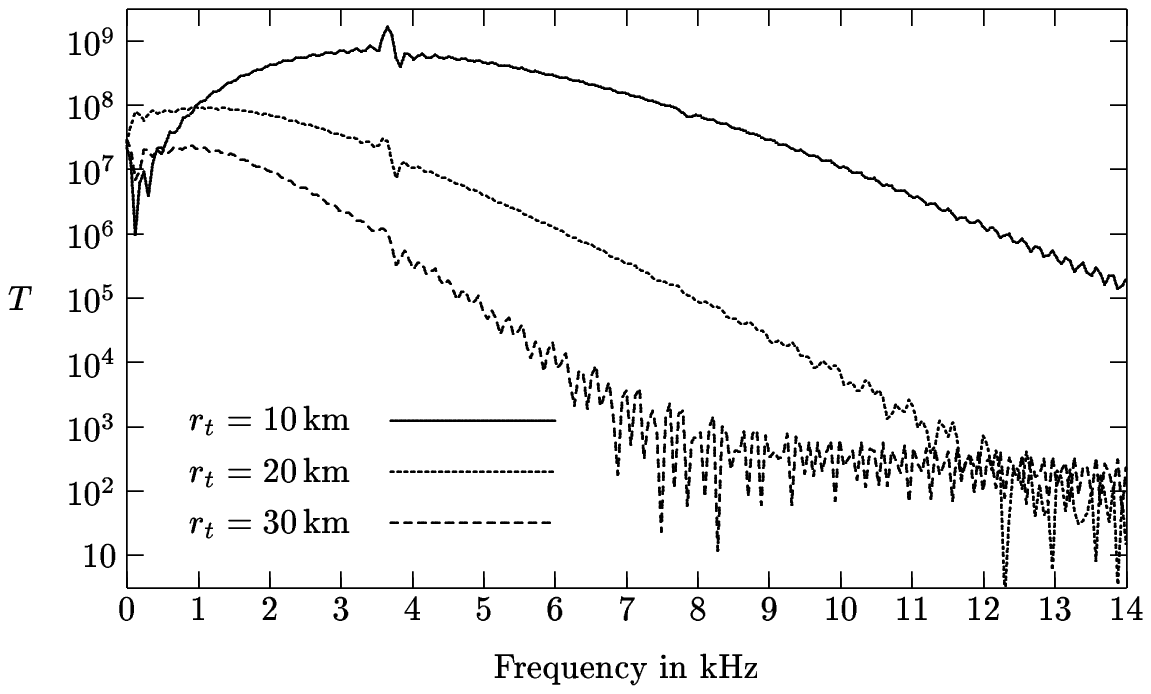}
\end{minipage}
\vspace*{3mm}
\caption{\label{fig:ST_v0=0.5fft}Fourier transform of the waveforms 
for $v_0 = -0.5$}
\end{figure}

\begin{figure}[hhh]
\leavevmode
\begin{minipage}{8.5cm}
\epsfxsize=\textwidth
\epsfbox{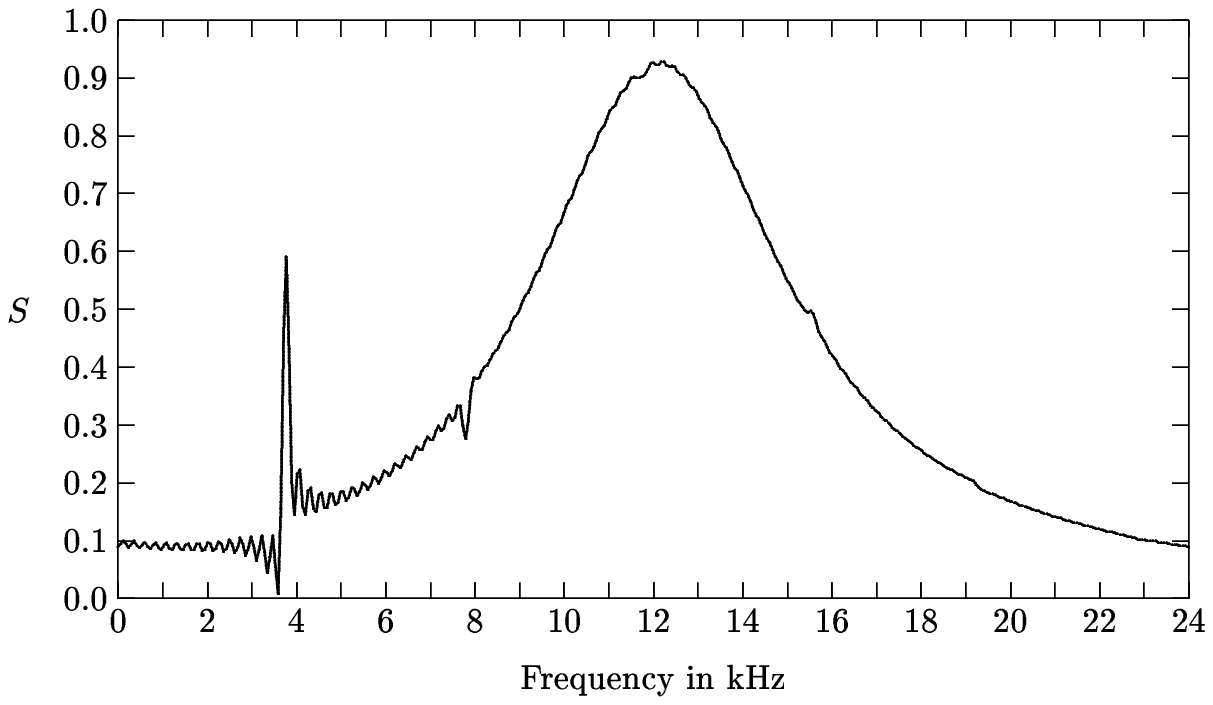}
\end{minipage}
\hspace*{2mm}
\begin{minipage}{8.5cm}
\epsfxsize=\textwidth
\epsfbox{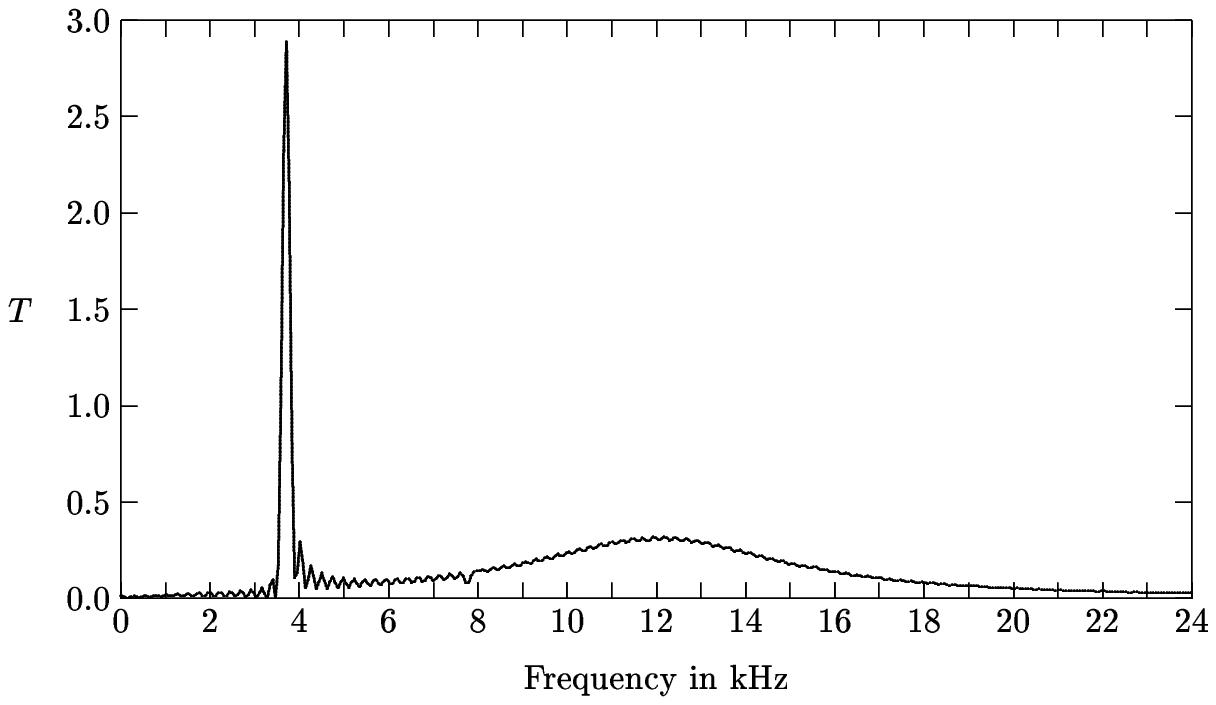}
\end{minipage}
\vspace*{3mm}
\caption{\label{fig:ST_v0=0.97fft}Fourier transform of the waveform for
$r_t = 10$km and $v_0 = -0.97$. The presence of the first $w$-mode is
clearly visible.}
\end{figure}

In Figs.~\ref{fig:ST_v0=0.1} through \ref{fig:ST_v0=0.97}, we show
the waveforms of $S$ and $T$ for different initial radial velocities
$v_0$ and different values of the radius of the turning point $r_t$.
In Figs.~\ref{fig:ST_v0=0.1fft} through \ref{fig:ST_v0=0.97fft} we
show the associated power spectra.

For all cases considered, the waveforms always have the same overall
features. The first part of the signal consists of an increase in the
amplitude that results of the particle entering the strong
gravitational field of the neutron star. After a few irregular
oscillations which correspond to the particle curving around the
neutron star, the final signal will be dominated by the quasi-normal
ringing of the neutron star, which is confirmed by Fourier analyzing
the signal.

For slow particles the power spectrum shows one broad peak around the
frequency that corresponds the the angular velocity of the particle at
the turning point $r_t$ and some little sharp peaks that correspond to
the fluid modes of the neutron star ($f$-mode and lowest $p$-modes).
From the waveforms it is clear that the excitations strengths of the
modes decrease rapidly as the turning point $r_t$ of the particle
moves farther away from the neutron star.

When the particle reaches the speed of light, the picture changes
drastically. For initial velocities $v_0 \lesssim -0.6$ the appearance
of the $w$-mode becomes evident. In the ultra-relativistic case of $v_0
= -0.97$ (Fig.~\ref{fig:ST_v0=0.97}) which corresponds to $E \approx
4.5$ it is clear that the lion's share of the energy gets radiated
through the first $w$-mode channel. Note, that in obtaining the power
spectrum of Fig.~\ref{fig:ST_v0=0.97fft}, we have cut off the initial
radiation and have taken the Fourier transformation right where the
$w$-mode ringing starts. It is interesting to see that even if the
particle's turning point is quite far away ($r_t = 50\,$km) the
$w$-mode is still visible.

All the calculations have been performed using a stellar model with
central energy density of $\eps_0 = 5.7\times 10^{15}\,$g$\,$cm$^{-3}$
which leads to a mass of 1.35 M$_\odot$ and radius of $7.52\,$km. This model
is right below the maximum mass that is allowed for the polytropic
equation of state with $\Gamma = 2$ and $\kappa = 100\,$km$^2$. It is
clear that the results will not change too much if we change the
equation of state, or if we consider models with different
masses. Only for very light neutron stars, the $w$-modes can become
invisible again. For ultra-compact stars, we expect the long-lived
trapped modes to be a dominant part of the power spectrum.

\section{Conclusions}
\label{sec:con}
We have performed simulations of particles on circular orbits as well
as on scattering trajectories. The circular orbits were done for code
testing reasons, and we found that the radiated energies agree
perfectly with those obtained by Cutler et al.~\cite{CFPS93} who
looked at a particle orbiting a black hole.

For the scattering orbit, we discussed the numerical difficulties
associated with an evolution code. However, the results are 
accurate enough to address the issue of the excitation of 
$w$-modes by orbiting particles. First of
all, it became clear that in order to excite a significant amount of
$w$-modes the particle's velocity at infinity must be $\gg 0.5$c.
Secondly, the excitation strength of
the modes rapidly decreases as the particle's turning point $r_t$
increases. This is in agreement with the results of Ferrari et
al.~\cite{FGB99}, who report that no significant $w$-mode
excitation is observable. However, they did not consider the case
where the particle has an energy $E\gg 1$, which is necessary if it
were to excite $w$-modes. This was done by Andrade and Price for the
odd-parity case \cite{AP99}. Here, too, they find excitations of
$w$-modes. 

The following question now arises: Can we infer anything about what
happens in an astrophysical event? Our results show that the
particle's initial velocity has to be incredibly high in order to
produce a significant amount of $w$-mode excitation. It is clear that
there do not exist any astrophysical events at all that could
accelerate the particle (which, of course, represents some heavy
extended object, like a planet or even another neutron star) up to
some significant fraction of the speed of light. Hence, we might
safely conclude that any astrophysical scenario that might be
simulated by a particle orbiting a neutron star will not produce any
detectable amount of $w$-modes.

The alternative is then to excite $w$-modes from the collision of the
particle with the neutron star. Our method could not be applied to
this situation, but we could speculate about the possible
implications. It has been suggested that in this case there might be
some significant excitation of $w$-modes. In particular, simulations
of a particle that spirals onto a ultra-compact constant density star
where done by Borelli \cite{Bor97} who found that, indeed, both the
trapped modes and the $w$-modes where excited, the latter much less,
however, than the former. One serious drawback of those calculations,
however, was that as soon as the particle had hit the surface of the
star, the calculation was stopped.  This lead to an overestimation of
the high frequency components in the power spectrum which, of course,
are represented by the $w$-modes. Since in this study, only odd parity
modes were investigated, it is clear that the $w$-modes will always
show up in the spectrum because they will not be screened by the
presence of fluid modes what can happen in the even parity case. Now,
even in the particle limit, it is not possible to simulate a collision
since it not clear at all what happens if the particle hits the
surface.

A first step, though, might be to let the particle go right through
the neutron star without being affected by the presence of the neutron
star matter. If this creates strong $w$-mode excitations even for
particles with low initial velocities, then one could conclude that,
in a realistic scenario where the physics of the impact is included,
there still might be $w$-modes present. Of course, in this case, it
seems reasonable to infer that at least the fluid modes would be much
more strongly excited than in a scattering event.



Furthermore, our results seem to indicate that the presence of the
$w$-modes in the signal is somehow related with the value of $S$. We
have seen that for ultra-relativistic particles the values of $rS$
greatly dominates the one of $T/r$. And it is only there where we find
a significant amount of $w$-modes. This particular role of $S$ seems
to be corroborated by the results of paper I, where it was shown that
the occurrence of $w$-modes strongly depends on the chosen initial
data. It was found that for conformally flat initial metric
perturbations (i.e.~$S_0 = 0$) the evolution shows practically no
$w$-modes at all. However, if one takes the same initial fluid
perturbation, sets $T_0 = 0$ and solves for $S$, suddenly one can
observe a quite large excitation of the $w$-modes in the signal. This
also agrees with results in \cite{AAKS98} and \cite{AAKLPR99}

Still, the question remains, as to what happens in a realistic
scenario. We have excluded the possibility of $w$-modes excitation by
means of a realistic scattering process. However, it is not possible
to relate our results to the merger process of a binary neutron star
system since one cannot adequately described the system within perturbation
theory. Only, if the final object does not immediately collapse to a
black hole, it might wildly oscillate and emit some significant
radiation through $w$-modes. It will only be through nonlinear 
evolutions that this issue might finally be settled.

\begin{acknowledgements}
Work supported in part by NSF grants PHY9800973 and PHY9800970.
J.R. was supported by a HSPIII grant from the DAAD (Germany).
\end{acknowledgements}

\begin{appendix}

\section{The source terms of the particle}

The particle's energy-momentum tensor \eqref{part_tmn} will be
expanded using the orthonormal set of tensor harmonics
$\left\{[\mz^A_{lm}]_{\mu\nu}\right\}_{A=1..10}$ given by Zerilli
\cite{Zer70a}:
\begin{align}
	\label{expan}
	{\cal T}_{\mu\nu} &= \sum_{l = 0}^{\infty}
	\sum_{m = -l}^l \sum_{A = 1}^{10}
	\hat{t}^{lm}_A [\mz^A_{lm}]_{\mu\nu}\;.
\end{align}
The coefficients $\hat{t}^{lm}_{A}$ can be obtained by means of the
orthogonality relation
\begin{align}\label{prod}
        \int_{S^2}[\mz^{A}_{lm}]^{*\mu\nu}
        [\mz^{A'}_{l'm'}]_{\mu\nu}\,{\rm d\Omega}
        &= s_A\delta_{AA'}\delta_{ll'}\delta_{mm'}\;,
\end{align}
where the asterisk denotes complex conjugation and
\begin{align}
        [\mz^{A}_{lm}]^{\mu\nu} &= \eta^{\mu\kappa}\eta^{\nu\sigma}\,
        [\mz^{A}_{lm}]_{\kappa\sigma}\;.
\end{align}
Because of the use of the inverse Minkowsky metric $\eta^{\mu\nu}$ to
raise the indices, the inner product (\ref{prod}) is not positive
definite and it is $s_A = -1$ for the harmonics with nonzero
$0j$-components ($A$ = 2, 3, 4).

By using the orthonormality condition \eqref{prod} we can compute the 
coefficients $\hat{t}^{lm}_A$ through
\begin{align}\label{coeff}
        \hat{t}_A^{lm} &= \int_{S^2}[\mz^A_{lm}]^{*\mu\nu}{\cal T}_{\mu\nu}
        \,{\rm d\Omega}\;.
\end{align}
We thus obtain the following set (\cite{Zer70c} with some corrections):
\begin{eqnarray*}
	\hspace*{-5cm}
	&\left[tt\right]& \quad\,
	\hat{t}_{1}^{lm} = e^{4\nu}\frac{\mu}{r^2}\frac{dt}{d\tau}
	\delta(r-R(t))Y^*_{lm}\\
	&\left[tr\right]& \quad\,
	\hat{t}_{2}^{lm} = \mbox{i}\sqrt{2}\frac{\mu}{r^2}\frac{dR}{d\tau}
	\delta(r-R(t))Y^*_{lm}\\
	&\left[\begin{array}{r}
	t\theta\\[1ex]
	t\phi
	\end{array}\right]&
	\left\{\begin{array}{l}\mbox{$\displaystyle
	\hat{t}_{3}^{lm} = e^{2\nu}\frac{2\mbox{i}\mu}{r\sqrt{2l(l+1)}}
	\delta(r-R(t))\frac{d}{d\tau}Y^*_{lm}$}\\
	\mbox{$\displaystyle
	\hat{t}_{4}^{lm} = e^{2\nu}\frac{2\mu}{r\sqrt{2l(l+1)}}
	\delta(r-R(t))\(\frac{1}{\sin\Theta}\frac{d\Theta}{d\tau}
	\df{}{\Phi} - \sin\Theta\frac{d\Phi}{d\tau}\df{}{\Theta}\)
	Y^*_{lm}$}
	\end{array}\right.\\
	&\left[rr\right] & \quad\,
	\hat{t}_{5}^{lm} = e^{4\lambda}\frac{\mu}{r^2}
	\frac{d\tau}{dt}\(\frac{dR}{d\tau}\)^2
	\delta(r-R(t))Y^*_{lm}\\
	&\left[\begin{array}{r}
	r\theta\\[1ex]
	r\phi
	\end{array}\right]
	& \left\{\begin{array}{l}\mbox{$\displaystyle
	\hat{t}_{6}^{lm} = e^{2\lambda}\frac{2\mu}{r\sqrt{2l(l+1)}}
	\frac{d\tau}{dt}\frac{dR}{d\tau}\delta(r-R(t))
	\frac{d}{d\tau}Y^*_{lm}$}\\
	\mbox{$\displaystyle
	\hat{t}_{7}^{lm} = e^{2\lambda}\frac{2\mbox{i}\mu}{r\sqrt{2l(l+1)}}
	\frac{d\tau}{dt}\frac{dR}{d\tau}\delta(r-R(t))
	\(\sin\Theta\frac{d\Phi}{d\tau}\df{}{\Theta}-
	\frac{1}{\sin\Theta}\frac{d\Theta}{d\tau}\df{}{\Phi}
	\)Y^*_{lm}$}
	\end{array}\right.\\
	&\left[\begin{array}{r}
	\theta\theta\\[1ex]
	\theta\phi\\[1ex]
	\phi\phi
	\end{array}\right]
	& \left\{\begin{array}{l}\mbox{$\displaystyle
	\hat{t}_{8}^{lm} = \frac{\mu}{\sqrt{2l(l+1)(l-1)(l+2)}}
	\frac{d\tau}{dt}\delta(r-R(t))
	\left[\(\(\frac{d\Theta}{d\tau}\)^2
	- \sin^2\Theta\(\frac{d\Phi}{d\tau}\)^2\)W^*_{lm}
	+ 2\frac{d\Theta}{d\tau}\frac{d\Phi}{d\tau}X^*_{lm}\right]$}\\
	\mbox{$\displaystyle
	\hat{t}_{9}^{lm} = \frac{\mu}{\sqrt{2}}\frac{d\tau}{dt}
	\delta(r-R(t))\(\(\frac{d\Theta}{d\tau}\)^2
	+ \sin^2\Theta\(\frac{d\Phi}{d\tau}\)^2\)
	Y^*_{lm}$}\\
	\mbox{$\displaystyle
	\hat{t}_{10}^{lm} = \frac{\mbox{i}\mu}{\sqrt{2l(l+1)(l-1)(l+2)}}
	\frac{d\tau}{dt}\delta(r-R(t))\sin\Theta
	\left[\(\(\frac{d\Phi}{d\tau}\)^2
	- \frac{1}{\sin^2\Theta}\(\frac{d\Theta}{d\tau}\)^2\)X^*_{lm}
	+ 2\frac{d\Theta}{d\tau}\frac{d\Phi}{d\tau}W^*_{lm}\right]$}
	\end{array}\right.\\
\end{eqnarray*}
with
\begin{align}
	\label{Xdef} X_{lm} &:= 2\(\df{}{\theta} - \cot\theta\)\df{}{\phi} Y_{lm}\\
	\label{Wdef} W_{lm} &:=
        \(l\(l+1\) + 2\,\dff{}{\theta}\)Y_{lm}\;.
\end{align}
Here, $Y^*_{lm}, W^*_{lm}$ and $X^*_{lm}$ are functions of the
particle's angular position $\Theta$ and $\Phi$ parametrized by the
coordinate time $t$, therefore all derivatives with respect to proper
time $\tau$ are to be understood as
\begin{align}
	\frac{d}{d\tau} = \frac{dt}{d\tau}\frac{d}{dt}
	= e^{2\lam}E\frac{d}{dt}\;.
\end{align}
Since the evolution equations are expanded using the Regge-Wheeler
harmonics, which do not form an orthonormal set, we have to convert
the above coefficients $\hat{t}_A^{lm}$ into the coefficients
$t_A^{lm}$ that would result from an expansion of the
energy-momentum tensor into Regge-Wheeler harmonics. Using the
relationship between the different sets of tensor harmonics given in
\cite{Zer70a} and \cite{Ruoff2000} we find:
\begin{align*}
	t_1^{lm} &= \hat{t}_{1}^{lm}\\
	t_2^{lm} &= \frac{\mbox{i}}{\sqrt{2}}\hat{t}_{2}^{lm}\\
	t_3^{lm} &= \frac{\mbox{i} r}{\sqrt{2l(l+1)}}\hat{t}_{3}^{lm}\\
	t_4^{lm} &= -\frac{r}{\sqrt{2l(l+1)}}\hat{t}_{4}^{lm}\\
	t_5^{lm} &= \hat{t}_{5}^{lm}\\
	t_6^{lm} &= \frac{r}{\sqrt{2l(l+1)}}\hat{t}_{6}^{lm}\\
	t_7^{lm} &= -\frac{\mbox{i} r}{\sqrt{2l(l+1)}}\hat{t}_{7}^{lm}\\
	t_8^{lm} &= \frac{2r^2}
	{\sqrt{2l(l+1)(l-1)(l+2)}}\hat{t}_{8}^{lm}\\
	t_9^{lm} &= \frac{r^2}{\sqrt{2}}\(\hat{t}_{9}^{lm} + \sqrt{\frac{
	l(l+1)}{(l-1)(l+2)}}\hat{t}_{8}^{lm}\)\\
	t_{10}^{lm} &= -\frac{2\mbox{i} r^2}
	{\sqrt{2l(l+1)(l-1)(l+2)}}\hat{t}_{10}^{lm}\;.
\end{align*}
We now restrict the motion of the particle to the equatorial plane
$\Theta = \frac{\pi}{2}$. In this case it is $\frac{d\Theta}{d\tau} =
0$ and $\sin\Theta = 1$, and we can use the geodesic equations
\eqref{geo1} and \eqref{geo3} to substitute all expressions containing
derivatives with respect to proper time $\tau$. Furthermore, we can
obtain quite simple relations for the derivatives of $Y^*_{lm}$:
\begin{align}
	\df{}{\Phi}Y^*_{lm} &= -\mbox{i} mY^*_{lm}\\
	\dff{}{\Theta}Y^*_{lm} &= \(m^2 - l(l+1)\)Y^*_{lm}\;.
\end{align}
This gives us a somewhat simpler set of coefficients:
\begin{align*}
	t_1^{lm} &= e^{2\nu}\frac{\mu E}{r^2}\delta(r-R(t))
	Y^*_{lm}\\
	t_2^{lm} &= -e^{2\lambda}\frac{\mu E}{r^2}
	v_r\delta(r-R(t))Y^*_{lm}\\
	t_3^{lm} &= e^{2\nu}\frac{\mbox{i} m\mu L}{r^2l(l+1)}\delta(r-R(t))
	Y^*_{lm}\\
	t_4^{lm} &= e^{2\nu}\frac{\mu L}{r^2l(l+1)}\delta(r-R(t))
	\df{}{\Theta}Y^*_{lm}\\
	t_5^{lm} &= e^{6\lambda}\frac{\mu E}{r^2}v^2_r\delta(r-R(t))
	Y^*_{lm}\\
	t_6^{lm} &= -e^{2\lambda}\frac{\mbox{i} m\mu L}{r^2l(l+1)}v_r
	\delta(r-R(t))Y^*_{lm}\\
	t_7^{lm} &= e^{2\lambda}\frac{\mu L}{r^2l(l+1)}v_r
	\delta(r-R(t))\df{}{\Theta}Y^*_{lm}\\
	t_8^{lm} &= e^{2\nu}\frac{\mu L^2(l(l+1) - 2m^2)}
	{r^2El(l+1)(l-1)(l+2)}\delta(r-R(t))Y^*_{lm}\\
	t_9^{lm} &= e^{2\nu}\frac{\mu L^2(l(l+1) - m^2 - 1)}{r^2E(l-1)(l+2)}
	\delta(r-R(t))Y^*_{lm}\\
	t_{10}^{lm} &= -e^{2\nu}\frac{2\mbox{i} m\mu L^2}
	{r^2El(l+1)(l-1)(l+2)}\delta(r-R(t))
	\df{}{\Theta}Y^*_{lm}\;,
\end{align*}
where $v_r = \frac{dR}{dt}$ is the radial velocity of the particle.
The field equations also require the computation of the trace 
\begin{align}
	\calt &= g^{\mu\nu}\calt_{\mu\nu} \;=\; \sum_{l,m}t^{lm}Y_{lm}\;,
\end{align}
with 
\begin{align}
	t^{lm} &= -e^{2\lambda}\,t_1^{lm}+ e^{-2\lambda}\,t_5^{lm}
	- \frac{l\(l+1\)}{r^2}\,t_8^{lm} + \frac{2}{r^2}\,t_9^{lm}\;.
\end{align}
Using the explicit forms of the coefficients we obtain
\begin{align}
	t^{lm} &= \frac{\mu}{r^2}\,\delta(r-R(t))
	\(E\,v^2_r\,e^{4\lambda} - E + e^{2\nu}\frac{L^2}{r^2 E}\)
	Y^*_{lm}\;,
\end{align}
and by making use of the geodesic equation (\ref{geo2}) we can reduce
this expression to
\begin{align}
        t^{lm} &= -e^{2\nu}\frac{\mu}{r^2 E}\,\delta(r-R(t))Y^*_{lm}\;.
\end{align}
\end{appendix}

\vspace*{5cm}

\newpage
\vspace*{-2cm}
\vspace*{-1cm}
\vspace*{-1cm}

\newpage

\vspace*{1cm}
\begin{figure}[hhh]
\leavevmode
\begin{minipage}{8.5cm}
\epsfxsize=\textwidth
\epsfbox{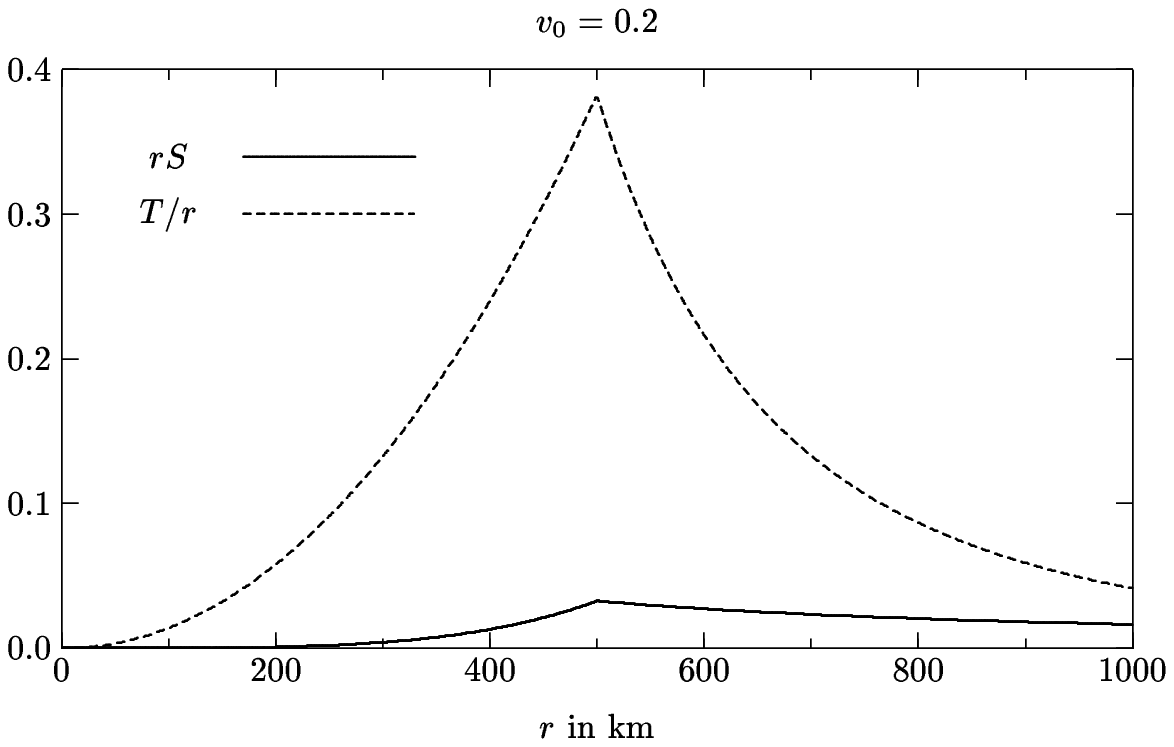}
\end{minipage}
\hspace*{2mm}
\begin{minipage}{8.7cm}
\epsfxsize=\textwidth
\epsfbox{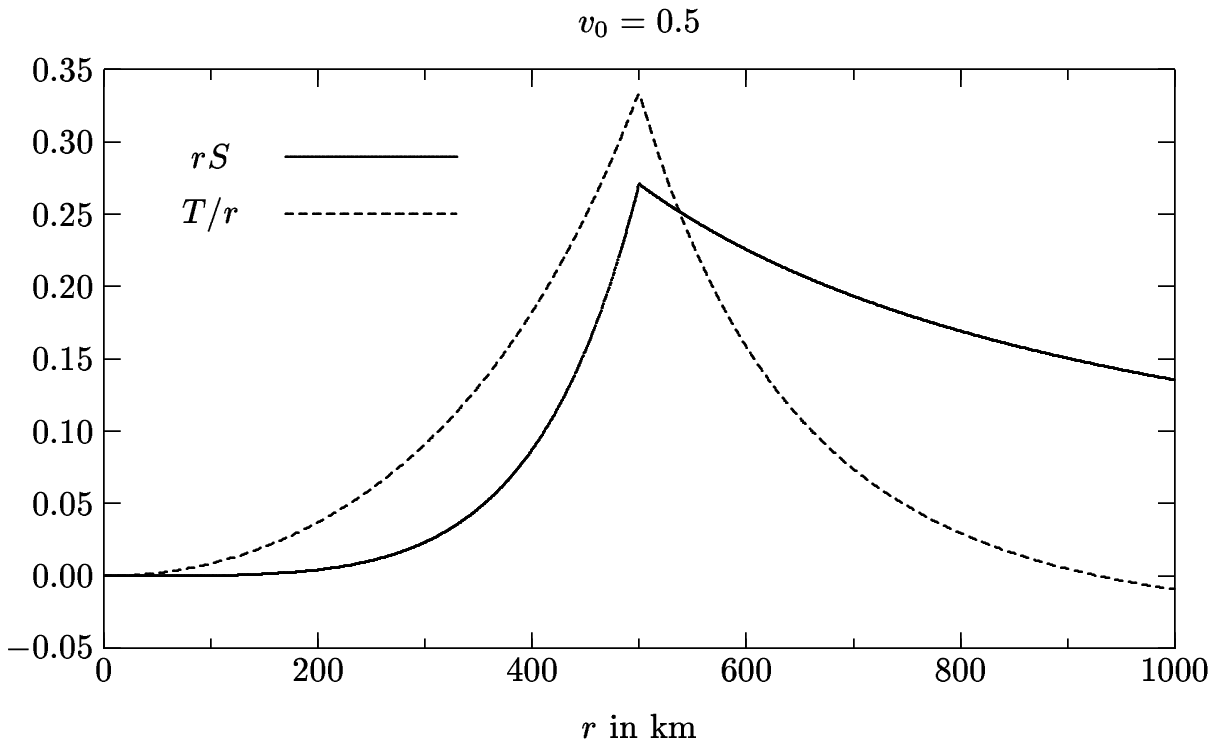}
\end{minipage}
\end{figure}
\vspace*{1cm}

\vspace*{2cm}

\newpage

\vspace*{1cm}

\vspace*{1cm}

\newpage

\vspace*{1cm}

\vspace*{1cm}

\end{document}